\documentclass[sigconf]{acmart}
\AtBeginDocument{%
  \providecommand\BibTeX{{%
    \normalfont B\kern-0.4em{\scshape i\kern-0.24em b}\kern-0.8em\TeX}}}

\setcopyright{acmcopyright}
\copyrightyear{}
\acmYear{}
\acmDOI{XXXXXXX.XXXXXXX}

\acmConference[Conference acronym 'XX]{}{}{}
\acmPrice{15.00}
\acmISBN{978-1-4503-XXXX-X/18/06}

\usepackage{bbding}
\usepackage{tcolorbox}
\usepackage{bigstrut,bigdelim,multirow}
\usepackage[figuresright]{rotating}
\usepackage{graphicx}

\begin{document}

\title{Neural Program Repair : Systems, Challenges and Solutions }


\author{Wenkang Zhong}
\affiliation{%
  \institution{State Key Laboratory for Novel Software Technology, Nanjing University}
  \city{Nanjing, Jiangsu}
  \country{China}}
\email{dg21320011@smail.nju.edu.cn}

\author{Chuanyi Li$^*$}
\affiliation{%
  \institution{State Key Laboratory for Novel Software Technology, Nanjing University}
  \city{Nanjing, Jiangsu}
  \country{China}
}
\email{lcy@nju.edu.cn}
\author{Jidong Ge}
\affiliation{%
  \institution{State Key Laboratory for Novel Software Technology, Nanjing University}
  \city{Nanjing, Jiangsu}
  \country{China}
}
\email{gjd@nju.edu.cn}
\author{Bin Luo}
\affiliation{%
  \institution{State Key Laboratory for Novel Software Technology, Nanjing University}
  \city{Nanjing, Jiangsu}
  \country{China}
}
\email{luobin@nju.edu.cn}

\renewcommand{\shortauthors}{}

\begin{abstract}
  Automated Program Repair (APR) aims to automatically fix bugs in the source code. Recently, with advances in Deep Learning (DL) field, there has been an increase of Neural Program Repair (NPR) studies that use neural networks to model the patch-generation process. NPR approaches have a significant benefit in applicability over prior APR techniques because they do not require any specifications (e.g., a test suite) when generating patches. For this reason, NPR has recently become a popular research topic. 
  In this paper, We undertake a literature review of latest NPR systems to help interested readers understand advancements in this emerging field. We begin by introducing background information of NPR. Next, to make the various NPR systems more understandable, we split them into a four-phase pipeline and discuss various design choices for each phase. To investigate the motivations of different design choices, We further highlight a number of challenges and summarize corresponding solutions adopted by existing NPR systems. Finally, we suggest some intriguing directions for the future research.
\end{abstract}



\maketitle

\section{Introduction}
Since debugging is a costly but necessary activity to ensure software quality \cite{quantify}, Automated Program Repair (APR) \cite{APR}, which aims to automatically fix bugs without human intervention, has become an important research topic in both software engineering and artificial intelligence communities. 
In the last decade, the majority of popular APR approaches are test-suite-based ones\cite{test-based:2011genprog,test-based:2013Semfix,test-based:2016Astor,test-based:2017ACS,test-based:2017ssFix,test-based:2018CapGen,test-based:2018Cardumen,test-based:2018SimFix,test-based:2020ARJA,test-based:2021JAID}, in which test cases of the buggy program are adopted as specifications to guide the repair process. During the bug-fixing process, candidate patches are continuously generated and applied to the buggy program until the patched program satisfies the expected behavior described by the specification. 
However, the applicability of test-suite-based APR systems is limited in practice, as test suites are difficult to come by and their quality is often inadequate for use as a repair specification \cite{phoenix}. 

\begin{figure}[t!]
  \centering
  \includegraphics[width=\linewidth]{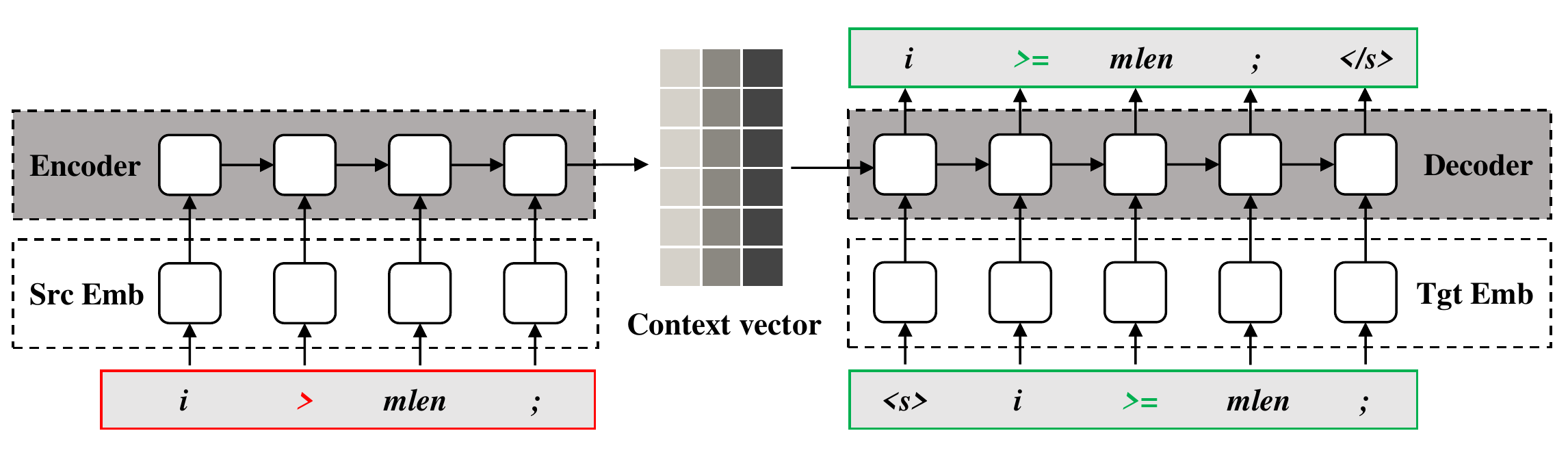}
  \caption{An example of NPR. The buggy statement is highlighted in red dotted box and the repaired statement in green.}
  \label{fig:example}
\end{figure}

Recently, researchers start paying more attention to Neural Program Repair (NPR) approaches \cite{Ratchet,Tufano19,hoppity20,CoCoNut20,codit20,DLFix20,transformer-edit20,CodeBert-finetune21,Tang21,SequenceR19,joint,Recoder21,TFix21,execution-based,transferlearning,deepfix,break-it-fix-it,DrRepair,DeepDebug} since they have shown great potential on automatically generating patches without test suites. 
\begin{table*}[t!]
  \centering
  \setlength{\abovecaptionskip}{0cm}
  \small
  \caption{Distribution of included NPR systems along with the publication channels}
    \begin{tabular}{p{28em}l}
    \toprule
    \multicolumn{1}{l}{\textbf{Publication Channel}} & \textbf{Studies} \\
    \midrule
    \multicolumn{1}{l}{IEEE International Conference on Software Engineering (ICSE)} & CURE \cite{Cure21}, DLFix \cite{DLFix20}, RewardRepair \cite{execution-based}\\
    \multicolumn{1}{l}{International Conference on Machine Learning (ICML)} & TFix \cite{TFix21}, DrRepair \cite{DrRepair}\\
    \multicolumn{1}{l}{IEEE Transactions on Software Engineering (TSE)}  & SequenceR \cite{SequenceR19}\\
    \multicolumn{1}{l}{International Conference on Learning Representations (ICLR)} & Hoppity \cite{hoppity20}, Vasic \cite{joint} \\
    \multicolumn{1}{l}{International Conference on Automated Software Engineering (ASE)} & PatchEdits \cite{transformer-edit20} \\
    International Symposium on Software Testing and Analysis (ISSTA) & CoCoNut \cite{CoCoNut20}\\
    ACM International Conference on Mining Software Repository (MSR) & CodeBert-finetune \cite{CodeBert-finetune21}\\
    \multicolumn{1}{l}{Association for Computational Linguistics (ACL)} & Tang \cite{Tang21} \\
    \multicolumn{1}{l}{ACM Transactions on Software Engineering Methodology (TOSEM)} & Tufano \cite{Tufano19}, CODIT \cite{codit20}\\
    The Association for the Advancement of Artificial Intelligence (AAAI) & DeepFix \cite{deepfix} \\
    Proceedings of Machine Learning Research (PMLR) & BIFI \cite{break-it-fix-it} \\
    ACM Joint European Software Engineering Conference and Symposium on the Foundations of Software Engineering (ESEC/FSE) & Recoder \cite{Recoder21} \\
    \bottomrule
    \end{tabular}%
  \label{tab:distribution}%
\end{table*}%
The APR task is formulated as a translation from defective code to correct code in the NPR approach. Generally, the core component of a NPR system consists of an encoder and a decoder \cite{NMT-iclr15}, both of which contain several layers of neural networks that can receive sequential inputs. Take Figure \ref{fig:example} as an example: when repairing, the encoder first embeds the source buggy program into a semantic context vector, which is called \
\textit{Encoding}. The decoder then performs a multi-step generation of fix scheme. It calculates the probability distribution over a target vocabulary using the context vector and previously generated tokens as inputs at each step. This phase is called \textit{Decoding}. The Encoder-Decoder architecture has been applied in many text-processing tasks, such as Machine Translation \cite{NMT-iclr15} and Text Summarization \cite{TextSummarization}. Compared with text-processing tasks, NPR systems need to deal with programming language which has many differences from natural language. Thus, a complete NPR procedure usually includes two additional modules to deal with the input and the output: \textit{Preprocessing} and \textit{Patch Ranking}. The first module prepares source codes into standard inputs that neural encoders can receive, while the second module ranks outputs generated by the decoder to form a candidate patch set.

NPR approaches can be applied to a wider range of scenarios from development to production cycle than test-suite-based APR approaches because they only require aligned bug-fix program pairs for training. Code repositories, such as GitHub, are a good place to look for such information. However, the complexity of the APR task and the numerous processing steps present challenges in designing an effective NPR system. For example, when it comes to \textit{Context Extraction} (a sub-phase of the \textit{Preprocessing} module), a larger context may contain more repair ingredients on the APR side, but it also means a longer input that is harder to learn by neural networks, necessitating changes to the Encoder or Decoder Architecture.
To make the best decisions among various design options, developers must have a thorough understanding of both the Deep Learning and the APR domains, which impedes the NPR direction's follow-up research. But, first and foremost, there is no NPR-specific review to wrap up design space and discuss potential threats. Regarding the related surveys, they either focus solely on test-based approaches \cite{efficiency} or provide a broad view on a limited scope of NPR approaches from the perspective of techniques \cite{survey-1}. 

In this work, We provide a seminal overview of recent studies to help understand important design points on NPR systems. Different from aforementioned related reviews, We focus on detailed methodology of NPR systems, including (1) their various design choices at each step of the bug-fixing process and (2) challenges that most NPR systems face and the existing solutions they adopt. Specifically, we go over NPR's design space in depth, breaking down the overall procedure of NPR systems into a series of modules. Then, we detailed analyze design choices of each NPR systems on each module, discussing such choices' design reasons, shortcomings and potential improvement schemes. In addition, We discuss the challenges faced by each module and the corresponding solutions, and sum up with some generalized conclusions using existing evidences. We believe that this way is particularly easy to understand for readers and can facilitate the development of new NPR approaches. 

The remainder of this paper is organized as follows. Section \ref{Sec:Pre} introduces the preliminary knowledge about this field, such as task formulation, datasets and evaluation metrics. Section \ref{Sec:approaches} explicate various design choices of NPR systems and analyze their performance. Then, we identify common challenges and discuss the impact of existing solutions to these challenges in Section \ref{sec:challenge}. Furthermore, we conclude and discuss some future research directions in Section \ref{sec:future}.

\section{Included Studies}
Our goal is to provide a review of detailed design choices of NPR systems.
To include a NPR system in our review, we first search four  regular databases (IEEE Xplorer\footnote{http://ieeexplore.ieee.org/Xplore/}, ACM digital library\footnote{http://portal.acm.org}, SpringerLink\footnote{http://www.springerlink.com/} and GoogleScholar\footnote{http://scholar.google.com/}) with keywords "neural/translation/learning program/bug repair/fix". Then, we use an APR-related website\footnote{http://program-repair.org} and the living review of APR \cite{monperrus2020living} to supplement the search results. Since we aim to focus on their concrete design choices on methodology in this paper, we filter out irrelevant papers and NPR researches that do not propose a new NPR system. Finally, 17 NPR systems are included in our study. The distribution of all selected NPR systems along with their publication channels are presented in Table \ref{tab:distribution}. Among the included NPR systems, DeepFix \cite{deepfix}, DrRepair \cite{DrRepair} and BIFI \cite{break-it-fix-it} focus on compilation errors while the other systems \cite{Cure21,CoCoNut20,execution-based,hoppity20,joint,Tang21,Tufano19,Recoder21,CodeBert-finetune21,codit20,SequenceR19,transformer-edit20,TFix21,DLFix20} are proposed to fix dynamic errors of programs.

\section{Dataset and Metrics}\label{Sec:Pre}
In this section, we first give a formulation of the NPR task and then introduce available popular datasets and evaluation metrics. 

\textbf{Task.} The goal of the APR task is to generate patches for a buggy program automatically. A complete repair process usually consists of three steps: \textit{Fault Localization}, \textit{Patch Generation} and \textit{Patch Validation}. First, the fault statement of the buggy program will be located by fault localization techniques \cite{survey:FL}. Second, various patch generation models generate a number of candidate patches. Candidates should be validated by certain protocols in order to obtain the correct patches that are acceptable to developers. The NPR approach mainly does the work for \textit{patch generation} step. For NPR systems, inputs should be fault-localized programs (method-level or line-level) within a single file. Their outputs are possible patches that address the bug. 

\begin{table}[t!]
  \centering
  \small
  \setlength{\abovecaptionskip}{0cm}
  \caption{Popular datasets for NPR approaches. "TS" denotes whether the dataset provides corresponding test suites.}
    \begin{tabular}{lllll}
    \toprule
    \textbf{Datasets} & \textbf{Language} & \textbf{Source} & \multicolumn{1}{l}{\textbf{Size}} & \multicolumn{1}{l}{\textbf{TS}} \\
    \midrule
    Defects4J & Java  & OS Projects & 835   & \multicolumn{1}{l}{\Checkmark} \\
    QuixBugs & Java, python & Competition & 80    & \multicolumn{1}{l}{\Checkmark} \\
    BFP(small) & Java  & OS Projects & 58,350 &  \\
    BFP(medium) & Java  & OS Projects & 65,454 &  \\
    CodRep & Java  & OS Projects & 58,069 &  \\
    Restricted Test & JavaScript  & OS Projects & 243,054 &  \\
    \bottomrule
    \end{tabular}%
  \label{tab:dataset}%
\end{table}%

\textbf{Datasets.} We list widely adopted datasets in NPR research in Table \ref{tab:dataset}. Essentially, each data instance in all of these datasets contains an aligned bug-fix pair derived from programming competition submissions or Open-Source (OS) project commits. Defects4J is built on top of the version control system used by operating systems. It provides both buggy and fixed program versions with corresponding test suites to make bugs reproducible. QuixBugs \cite{dataset:QuixBugs} consists of 40 programs translated to both Python and Java, each with a bug on a single line. Owing to the test suites, these two also serve as the most popular evaluation datasets for all APR approaches. However, their sizes are relatively too small to train a learning-based NPR approach, with Defects4J (latest version) having 835 instances and QuixBugs having only 80 (40 of Java and 40 of Python). 
As a result, several larger datasets \cite{Tufano19,dataset:CodRep,TFix21} are constructed to allow for the training and evaluation of NPR systems, as shown in Table~\ref{tab:dataset}. Specifically, BFP \cite{Tufano19} provides two datasets for small methods (less than 50 tokens) and medium methods (50 to 100 tokens) extracted from commits between March 2011 and October 2017 on GitHub, while CodRep \cite{dataset:CodRep} mines bugs mined by several previous studies \cite{source1,source2,source3,source4,source5}. Similarly, the JavaScript bug-fix dataset Restricted Text \cite{TFix21} is constructed from 5.5 million GitHub commits.
In consideration of cost on collecting test suites, these larger datasets only provide human-written patches for validation.

\textbf{Evaluation Metrics.} The NPR approach usually predicts patches with the top confidence score to form the candidate set. If a candidate can pass all the given test case, it is regarded as a \textbf{plausible} patch. Such patches may not be accepted by developers for some reasons (i.e., introduce other faults) so further manual check is performed to ensure that they are \textbf{correct}. In this way, performances of NPR approaches can be measured as the number of \textbf{correct}$\backslash$\textbf{plausible} patches. And when in the absence of a corresponding test suite, candidates that are exactly the same with the human-written patches can be regarded as correct. \textbf{Top-K accuracy} is the evaluation metric in this case, where K is the size of the candidate set for each bug.

\section{NPR Systems}\label{Sec:approaches}
In this section, we review the existing NPR systems from the perspective of their design choices. First, we introduce the overall procedure of NPR systems in Section \ref{procedure}. Then, design choices on each module of the procedure are explicated in Section \ref{preprocessing} to \ref{sec:rank}. Finally, we give a summary of included studies to retrieve the SOTA systems and point out the limitation. 
\begin{figure*}[t!]
  \centering
  \includegraphics[width=0.80\textwidth]{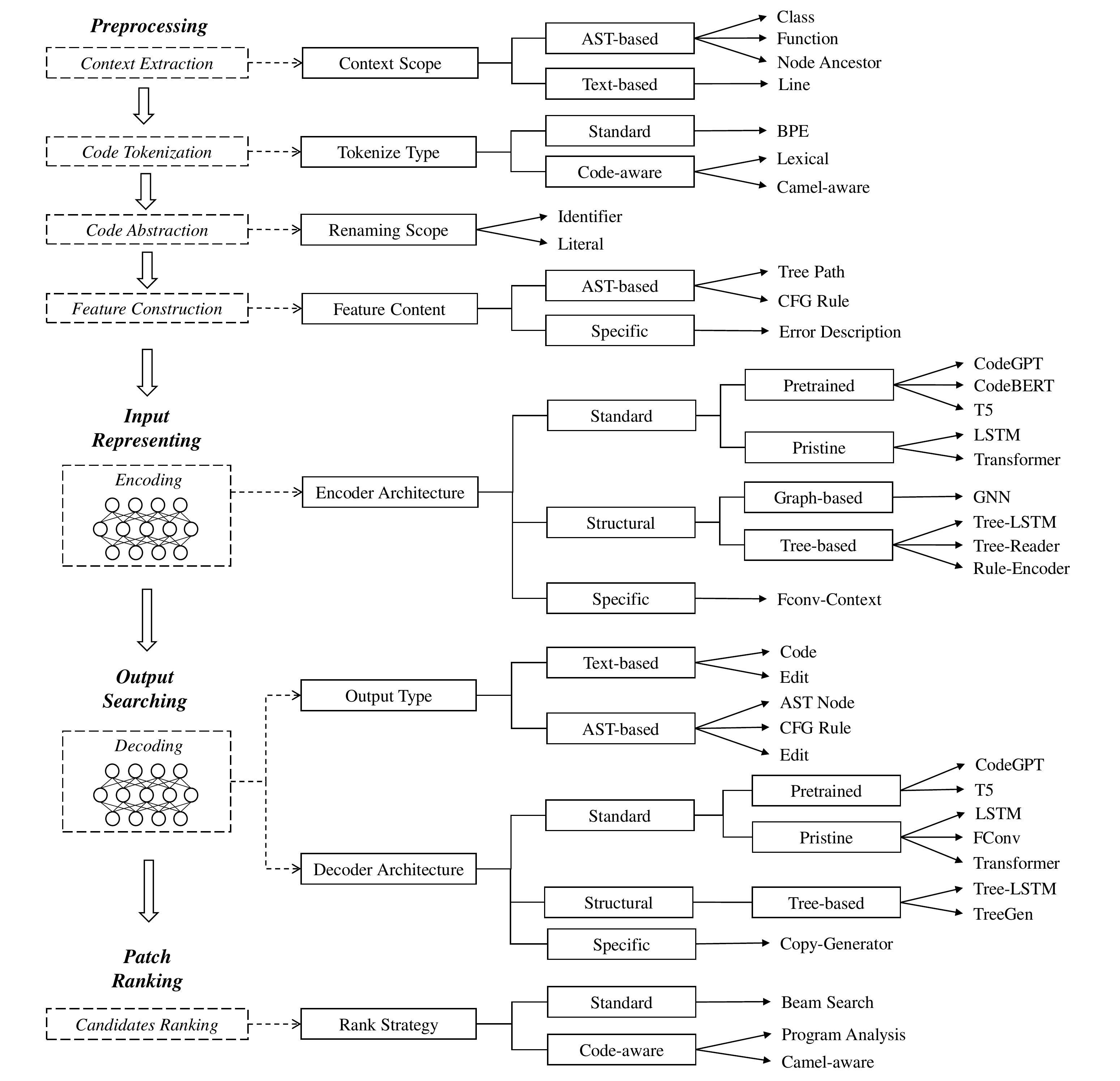}
  \caption{The overall procedure and corresponding design space of the NPR. The left part of the figure describes the processing pipeline and the right part categorized design choices of each module. }
  \label{fig:designchoices}
\end{figure*}
\subsection{Overall Procedure}\label{procedure}
Generally, NPR systems follow the processing progress illustrated in Figure \ref{fig:designchoices}. An end-to-end repair procedure consists of the following phases: 

(1) \textbf{\textit{Preprocessing}.} First, buggy programs are processed into sequential or structural forms that are acceptable by Neural Networks (NN) during the \textit{Preprocessing} phase. This phase has four sub-phases. In the \textit{Context Extraction} sub-phase, context surrounding the buggy fragment will be extracted to build a contextual input. Next, text-form codes need to be divided into lists of tokens through the \textit{Code Tokenization} sub-phase. There is an option for an \textit{Code Abstraction} sub-phase to simplify the input by renaming natural elements in the source code. Additional features such like the Abstract Syntax Tree (AST) will be constructed during the \textit{Feature Construction} sub-phase.

(2) \textbf{\textit{Input Representing}.} Next, processed inputs are fed into the encoder, which consists of several neural layers. The embedding layers of the encoder first represent the input as vectors, with tokens occurring in similar contexts having a similar vector representation. The next layers perform vector-based computation to produce a final contextual vector that contain rich semantics of the buggy program. In a sense, the encoder represents the knowledge of the programming language.

(3) \textbf{\textit{Output Searching}.} Then, the decoder module performs a multi-step fix generation, which is called \textit{decoding}. The decoder has a structure similar to the encoder. At each step, it takes the contextual vector produced by the encoder along with previously generated tokens as inputs and outputs the probability distribution over the target vocabulary. The fix can be code text or edit operations on the buggy program. This phase produced all potential patch candidates within the search space.

(4) \textbf{\textit{Patch Ranking}.} Finally, a rank strategy is necessary to reduce the patch candidates set to a reasonable size, considering the efficiency of subsequent patch validation. 

Each phase of the NPR procedure can be regarded as an independent function module and has positional design points. The following part will explicate various design choices of existing NPR systems on each module.

\subsection{Preprocessing}\label{preprocessing}
\subsubsection{Context Extraction}
For NPR task, the unprocessed input is a buggy file with fault location (usually at line level). Since a no-context line-to-line repair will gain a low performance \cite{SequenceR19}, most approaches will input the buggy line along with its surrounding context. It is important to decide the \textbf{context scope}. A wider context may contain more repair ingredients, but also introduces noise that degrades the performance of repair models. The commonly used approaches to extract context can be categorised as \textit{Text-based} or \textit{AST-based}.

A text-based extraction only considers context that is location-relevant to the buggy statement. The simplest way is to take a single buggy line which contains the buggy statement as context \cite{transformer-edit20}. Another case extracts the error context consisting of the buggy line and the tow neighboring lines \cite{TFix21}.
The more common approach to extract context is AST-based. Most studies select the nearest \textit{MethodDeclaration-type} ancestor of the buggy node as a root of contextual AST \cite{Tufano19,DLFix20,CoCoNut20,hoppity20,Recoder21,Tang21}. There are also some approaches \cite{codit20,CodeBert-finetune21} trying to make the model learn to fix bugs within a more narrow context -- the least common ancestor of the buggy node and the fix node. For a widest consideration, SequenceR \cite{SequenceR19} builds a class-level context with a length limit of 1,000 tokens. It keeps all the instance variables and initializers in the buggy class along with the signature of the constructor and non-buggy methods even if they are not called in the buggy method. 

\textit{Discussion and Insight.} How much context is enough for NPR systems?  A too long context may lead to the performance degradation of the model. First, the context may contain codes unrelated to the bug. Second, as known, neural network models are poor at dealing with long inputs. Therefore, to better use the context, researchers should focus on including much information that can be useful to fix the bug in a context as short as possible. 

\subsubsection{Code Tokenization}
Neural models can only receive sequential or structural formed input, so the textual source code needs to be divided into a list of tokens first, which is called \textit{tokenization}. For NPR task, the \textbf{Tokenize Type} can be \textit{Standard} ways that are popular in other field or being \textit{Code-aware} for maintaining semantics of the program.

For most natural languages, a tokenization way that fits human intuition is word-piece.
However, for tasks that use a large vocabulary, a word-piece tokenization will decrease the performance since neural models are pool at dealing with unseen words that are Out-Of-Vocabulary (OOV) \cite{OOV}. Thus, modern models usually adopt a Byte-Pair-Encoding \cite{BPE} tokenization, encoding rare and unknown words as subword units to mitigate the OOV problem. Tracing back to NPR task, BPE-based tokenization are also used in some NPR systems \cite{transformer-edit20,Cure21,CodeBert-finetune21} for the reason that programming languages use a more irregular vocabulary.

Programming need to follow predefined lexical rules. Before execution, they need to be decomposed into individual units by the lexical parser. Similarly, a code-aware tokenization that most approaches \cite{Tufano19,DLFix20,codit20,hoppity20,SequenceR19,TFix21,Recoder21,Tang21} take is to fed source codes into a language-specific lexer. Besides, some common naming rule will be considered during tokenization to reduce the number of uncommon tokens. For instance, CoCoNut \cite{CoCoNut20} separates variable and method names with a special "CaMel" token, using camel letters, underscores, and numbers to split a long identifier. 

\textit{Discussion and Insight.} Which type of tokenization is more suitable for the NPR task? A lexical tokenization is a safe choice to retain the semantics of source codes. However, as mentioned above, NPR systems must consider the OOV problem. The typical solution to the OOV problem is to use a BPE tokenization. However, BPE will break the semantics of source codes. For example, an identifier may be divided into several sub-tokens when BPE is used to tokenize source codes. Is it possible to use a lexical tokenization, meanwhile mitigating the OOV problem? This is a valuable question to be explored by follow-up researchers.

\subsubsection{Code Abstraction}
The purpose of code abstraction is to sharp the size of the vocabulary that NPR models use. The model generates sequential tokens by computing probability distributions over a predefined token vocabulary. When dealing with a large vocabulary composed of many possible output tokens, it may become inefficient or imprecise. Code abstraction aims to simplify the input program by renaming natural elements like identifiers and literals, and there are different choices among the \textbf{renaming scope}.

The first approach \cite{Tufano19} adopts an ID-replace strategy. At the beginning, the source code is divided into a  stream of tokens by a lexer. Then the strategy substitutes each identifier/literal within the tokenized stream with a unique ordinal ID. Some frequently appeared identifiers and literals like "add" or "replace" will be retained, since they represent common concepts. This abstraction way can significantly reduce the size of the vocabulary, but introduces some obvious drawbacks, i.e., the abstract patch that contains an ID not appeared in the source can't be concretized. A more modest option is to abstract literals only \cite{CoCoNut20,Cure21}, considering that most uncommon words in the vocabulary are brought by strings. DLFix \cite{DLFix20} implemented a renaming abstraction in order to increase the chance for model to learn fix in similar scenarios.  They keep the type of variable in the new name along with the invoked method. At present, abstract literals and infrequent identifiers in the source code is the best practice. But it is far from perfect because such abstraction will decrease the applicability of models to real-world scenarios.

\textit{Discussion and Insight.} According to a recent study \cite{controlled-experiment}, code abstraction is an efficient trick to improve the performance of NPR systems. Abstracted codes reduce the size of the vocabulary, which sharps the search space of NPR systems. However, in some cases, such abstraction will decrease the recall rates of NPR systems. For example, abstraction of strings and numbers can make some bugs irreparable because they are caused by wrong strings or numbers.

\subsubsection{Feature Construction}
Programs written in high-level programming languages have richer information than natural language texts. For example, a syntactically correct program can be parsed into an Abstract Syntax Tree (AST) that reflects structural syntactic information. Models may get additional repair ingredients or learn to use grammar rules from such features. 

Most features are constructed from an AST. Since a tree-structure AST is not suitable for normal neural models that require sequential inputs, an extra process for representing AST is needed. A common solution is to use the sequential traverse results of AST. In this way, some approaches \cite{codit20,Tang21} try to represent the AST with a sequence of Context-Free-Grammar (CFG) rules \cite{CFG-rules}. A CFG is a tuple $G = (N, \sum, P, S)$, where $N$ is a set of non-terminals, $\sum$ is a set of terminals, $P$ is a set of production rules and S is the start symbol. The AST that belongs to the language defined by $G$ can be parsed by applying the appropriate derivation rules from the start symbol $S$. Each CFG rule is regarded as a token of the vocabulary. For a non-traverse representation, DLFix \cite{DLFix20} generates four ASTs for each bug-fix pair: a tree for buggy method, a tree for fix method, a buggy subtree for buggy nodes and a fix subtree for fix nodes. The authors adopt a DL-based code summarization model \cite{CodeSummarization} to represent a subtree into a single vector. Besides, authors of Recoder \cite{Recoder21} treat the AST as a directional graph and embed it with an adjacent matrix.
The latest novel idea of feature construction is to get the results of program analysis tools.  TFix \cite{TFix21} uses the bug report produced by a static bug detector as an additional feature, which describes the location and the type of the bug.

\textit{Discussion and Insight.} The purpose of constructing extra features is to ease the learning phase of NPR system. Features such as the AST may contain intuitive information that are useful for the system to fix the bug. However, the concrete role of these features in fixing bugs is still unknown. A detailed exploration of why these features work can provide a useful guide to designing useful features for NPR.
\subsection{Input Representing}
The purpose of \textit{Input Representing} is to make the model understand the semantics of buggy programs with an encoder module, which is also called \textit{Encoding}. To this aim, the input will be mapped into a special semantic vector space by the encoder module, which is composed of multi-layer neural networks. In this phase, to design a proper \textbf{encoder architecture} is the most important. Architectures of encoders that existing NPR approaches use can be categorized as \textit{Standard}, \textit{Struture-aware}.

Standard encoders refer to those popular in the Natural Language Processing (NLP) field. Under this category, classic neural networks like Long Short-Term Memory (LSTM) architecture \cite{LSTM96} and Transformer \cite{transformer} are most commonly used \cite{Tufano19,codit20,transformer-edit20,SequenceR19,deepfix}. Another popular choice is to reuse models that have already trained on a large corpus of codes, which is called \textit{pre-train} \cite{pretrain-BERT}. Given the effectiveness of language models in the NLP domain, CURE \cite{Cure21} proposes to add a GPT \cite{GPT} module pre-trained on software code to a Neural Machine Translation (NMT) architecture. TFix \cite{TFix21} and RewardRepair \cite{execution-based} leverage T5, a Transformer-based \cite{transformer} model pre-trained on NLP tasks and fine-tune it for the task of generating code fixes. And the CodeBERT \cite{CodeBERT} model, a bimodal pre-trained language model for both natural and programming languages, is also used to adapt the APR task \cite{CodeBert-finetune21}.

Structure-aware encoders are used to capture the AST-based features. For an original tree-structure AST, DLFix \cite{DLFix20} encodes it with a Tree-LSTM \cite{Tree-LSTM} and Hoppity \cite{hoppity20} adopts a Gated Graph Neural Network (GGNN) \cite{GGNN}, treating the AST as a graph. For traverse results of ASTs, Tang \cite{Tang21} designs a Cross-Attention mechanism to make full use of token information and syntax information interactively. Recoder \cite{Recoder21} uses a special encoder called Code Reader that combines traverse results and AST-based graph through three sub neural layers.

Specifically, CoCoNut \cite{CoCoNut20} represent the buggy line and the context separately as a second encoder, independently of the chosen context. They believe this way can help the model learn useful relations from the context (such as potential donor code, variables in scope, etc.) without adding noise to the relations learned from the buggy line. 

\textit{Discussion and Insight.} A standard encoding way is to make input suitable for the encoder. On contrast, a structure-aware way modifies the encoder to receive structural features. The former is quite easy to implement, while the latter requires researchers to have some domain knowledge on both the deep learning and the program repair field.
\subsection{Output Searching}
\begin{table*}[htb]
\setlength{\abovecaptionskip}{0cm}
  \centering
    \caption{A summary of performances of advanced NPR systems and their design choices on each module. Only NPR systems that are evaluated on datasets which have been used at least twice are included. The best results for each dataset are highlighted in bold. Accuracy at @top-50,@top-5,@top-1 on CodRep, Restricted Text and BFP(small). }
  \small
    \begin{tabular}{|c|p{4em}|p{4.14em}|p{4em}|p{3.574em}|p{4.0em}|p{4.00em}|p{3.834em}|p{4em}|p{3.2em}|p{3.94em}|}
    \hline
     &  & \textbf{CoCoNut} & \textbf{Codit} & \textbf{Cure} & \textbf{DLFix} & \textbf{Recoder} & \textbf{Tufano} & \textbf{SequenceR} & \textbf{TFix} & \textbf{Tang}\\
    \hline
    \multicolumn{1}{|c|}{\multirow{8}[16]{*}{\rotatebox{90}{\textbf{Design Choices}}}} & \textbf{Context Scope} & Method & Node-Ancestor & Method & Method & Method & Method & Class & Line & Method \\
\cline{2-11}          & \textbf{Renaming Scope} & Literal & \textbackslash{} & Literal & Literal & Identifier & Identifier + Literal & \textbackslash{} & \textbackslash{} & Identifier +Literal \bigstrut\\
\cline{2-11}          & \textbf{Tokenize Type} & Camel-aware & Lexical & Camel-aware+BPE& Lexical & Lexical & Lexical & Lexical & BPE & lexical \bigstrut\\
\cline{2-11}          & \textbf{Features} & Code & Code +Rule & Code & AST   & Code+AST & Code  & Code & Code & Code+Rule \bigstrut\\
\cline{2-11}          & \textbf{Encoder  Architecture\bigstrut} & \multicolumn{1}{p{4em}|}{FConv-Context} & biLSTM & \multicolumn{1}{p{4.04em}|}{PT-GPT +FConv-Context} & Tree-LSTM & Code+AST +Path Readers & biLSTM & \multicolumn{1}{p{5.04em}|}{biLSTM} & \multicolumn{1}{p{4em}|}{PT-T5} & Token+ Grammar Emb \bigstrut\\
\cline{2-11}          & \textbf{Decoder Architecture\bigstrut} & \multicolumn{1}{p{4em}|}{Fconv} & biLSTM +copy & \multicolumn{1}{p{4.04em}|}{PT-GPT +Fconv} & Tree-LSTM & Edit-Decoder & biLSTM & \multicolumn{1}{p{5.04em}|}{biLSTM +copy} & \multicolumn{1}{p{4em}|}{PT-T5} & Grammar-Decoder  \bigstrut\\
\cline{2-11}          & \textbf{Output Type} & \multicolumn{1}{p{4em}|}{Code} & Code +Rule & \multicolumn{1}{p{4.04em}|}{Code} & Node  & Node Edit & Code  & \multicolumn{1}{p{5.04em}|}{Code} & \multicolumn{1}{p{4em}|}{Code} & Node \bigstrut\\
\cline{2-11}          & \textbf{Rank Strategy \bigstrut} & \multicolumn{1}{p{4em}|}{Beam Search} & Beam Search & \multicolumn{1}{p{4.04em}|}{Code-aware } & DL-based CLS & Beam      Search & Beam Search & \multicolumn{1}{p{5.04em}|}{Beam  Search} & \multicolumn{1}{p{4em}|}{Beam Search} & Beam Search \bigstrut\\
    \hline
    \multicolumn{1}{|c|}{\multirow{5}[10]{*}{\rotatebox{90}{\textbf{Performance}}}} & \textbf{Defects4J} & \multicolumn{1}{p{4em}|}{44/85} & 30/51 & \multicolumn{1}{p{4.04em}|}{57/104} & 30/65 & \textbf{71/x} & \textbackslash{} & \multicolumn{1}{p{5.04em}|}{18/61} & \multicolumn{1}{p{4em}|}{\textbackslash{}} & \textbackslash{} \bigstrut\\
\cline{2-11}          & \multicolumn{1}{l|}{\textbf{QuixBugs}} & \multicolumn{1}{p{4em}|}{13/20} & \textbackslash{} & \multicolumn{1}{p{4.04em}|}{\textbf{26/35}} & \textbackslash{} & 17/17 & \textbackslash{} & \multicolumn{1}{p{5.04em}|}{\textbackslash{}} & \multicolumn{1}{p{4em}|}{\textbackslash{}} & \textbackslash{} \bigstrut\\
\cline{2-11}          & \multicolumn{1}{l|}{\textbf{BFP(small)}} & \multicolumn{1}{p{4em}|}{\textbackslash{}} & \textbackslash{} & \multicolumn{1}{p{4.04em}|}{\textbackslash{}} & \textbackslash{} & \textbackslash{} & \multicolumn{1}{l|}{9.22\%} & \multicolumn{1}{p{5.04em}|}{\textbackslash{}} & \multicolumn{1}{p{4em}|}{\textbackslash{}} & \multicolumn{1}{l|}{\textbf{11.47\%}} \bigstrut\\
\cline{2-11}          & \multicolumn{1}{l|}{\textbf{CodRep}} & \multicolumn{1}{p{4em}|}{\textbackslash{}} & \textbackslash{} & \multicolumn{1}{p{4.04em}|}{\textbackslash{}} & \textbackslash{} & \textbackslash{} & \multicolumn{1}{l|}{14.050\%} & \textbf{30.80\%} & \multicolumn{1}{p{4em}|}{\textbackslash{}} & \textbackslash{} \bigstrut\\
\cline{2-11}          & \textbf{Restricted Text} & 16.40\% & \textbackslash{} &   \textbackslash{}   & \textbackslash{} & \textbackslash{} & \textbackslash{} & 23.60\% & \textbf{46.30\%} & \textbackslash{} \bigstrut\\
    \hline
    \end{tabular}%
  \label{tab:performance}%
\end{table*}%
\subsubsection{Output Type}
The output space is defined by the vocabulary that consists of tokens. During the \textit{Output Searching} phase, the decoder will make a probable search through the output space and calculates probability distributions over the target vocabulary. To model the generation of patches, the \textbf{output type} can be either \textit{AST-based} or \textit{Text-based}.

For textual generation, the most common way is to treat fixing as a token-to-token translation \cite{Tufano19,CoCoNut20,SequenceR19,Cure21,CodeBert-finetune21}. At each iteration, a textual code token will be outputted. Such design allows the decoder to generate the patch code directly. Specifically, Edits \cite{transformer-edit20} outputs edit operations such as \textit{insert} and \textit{delete} on the buggy program instead of code tokens, stemming from human's edit operations on fixing error text.

For AST-based generation, the model will first build an AST as the backstone of the potential fix. At each step of generation, the output can be a new node to expand the AST \cite{DLFix20}, or a modification of the buggy node \cite{hoppity20,Recoder21}. Another option is to generate CFG rules of the AST \cite{codit20,Tang21} first and then convert them to concrete programs. The AST-based way can ensure the syntactic correctness of generated codes to some extent.

\textit{Discussion and Insight.} The AST-based way models the output as a sequence of CFG rules, thus it can ensure the output codes' syntax correctness. However, for the same program, the number of CFG rules can be much bigger than individual code tokens. It means that an AST-based generation can be less precise, since the model needs to make more predictions to form a complete AST. A mitigation is to design a more simple form to represent the AST.
\subsubsection{Decoder Architecture}
The purpose of \textit{decoding} is to estimate the probability distribution of potential fix schemes. This is a multi-step process. At each step, the decoder outputs a conditional probability distribution over vocabulary, considering previously generated tokens and the context vector generated by the encoder. The \textbf{decoder architecture} represents the activity of generating patches. They can be categorized into \textit{Standard} or \textit{Structure-aware}.

Similar to standard encoders, standard decoders can also be \textit{pristine} or \textit{pretrained}. Standard architectures like LSTM \cite{LSTM96} and Transformer \cite{transformer} are also popular choices for NPR systems \cite{Tufano19,codit20,CoCoNut20,SequenceR19}.The only change is that some pre-training models can only be used to initialize the encoder \cite{CodeBert-finetune21}.

Some structure-aware decoders are tree-based since they model the decoding phase as a modification of the AST rather than generating textual codes. To this aim, DLFix \cite{DLFix20} adopts a Tree-LSTM \cite{Tree-LSTM} to generate a new node to replace the buggy node. And Recoder \cite{Recoder21} developed a syntactic edit decoder based on TreeGen \cite{TreeGen} to produce edit operations on the AST of buggy program. 

The APR task is different from translation in that only part of the words need to be changed. Considering this difference, some approaches \cite{codit20,transformer-edit20,SequenceR19} add a copy mechanism \cite{copy} to make the decoder task-aware. The copy mechanism allows a direct copy of tokens from the source program during decoding. 

\textit{Discussion and Insight.} We observe that researchers prefer to use standard decoders. However, such decoders are designed to model natural languages. When applied to the program repair task, they can even not ensure the syntactic correctness of the generated codes. Also, we notice that DLFix \cite{DLFix20} and Recoder \cite{Recoder21} use a structure-aware decoder to model the AST rather than textual codes, thus ensuring the syntactic correctness of outputs.  

\subsection{Patch Ranking}\label{sec:rank}

The number of potential patches that the decoder produced will be $S_v^l$ where $S_v$ is the size of vocabulary and $l$ is the max length of output. It is unrealistic to validate every candidate since they need be manually checked eventually. Therefore, A \textbf{rank strategy} is necessary in considering that too many candidates will significantly decrease the inference efficiency of the model and burden the effort on patch validation. 

Beam Search is a general search strategy in the DL domain, and it is also used as a patch ranking strategy by a lot of APR approaches \cite{Tufano19,codit20,DLFix20,SequenceR19,Tang21}. For each iteration during the generation, the beam search algorithm checks the $t$ most likely tokens ($t$ corresponds to the beam size) and ranks them by the total likelihood score of the next $s$ prediction steps ($s$ correspond to the search depth). In the end, the top $t$ most likely sequences ordered by the likelihood are kept. To filter candidates that are of low quality, Cure \cite{Cure21} adopts a code-aware beam search strategy, performing a Valid-Identifier-Check. The strategy first uses static analysis to get valid identifiers in the source program. Then during search, the probability of the generated identifier which is not valid in the source will be set to $-inf$. DLFix \cite{DLFix20} uses a more complex patch ranking strategy. They first derive the possible candidates with a set of program analysis filters and then adopt a DL-based classification to re-rank candidates.

\textit{Discussion and Insight.} Using a simple probability sampling to limit the number of candidates is not enough, since this strategy can even not ensure the syntactic correctness of candidates. More code-aware filters should be developed to perform a reliable rank. For example, researchers can design a code-related penalty for the rank strategy. If a candidate contain some primary errors such as the compilation error, the rank of it should be lower.
\subsection{The State-Of-The-Art}

To summary, we provide a table view of NPR systems that achieve remarkable performance. These systems with their design choices on each module are summarized in Table \ref{tab:performance}. Besides, We list their performances on five datasets for comparison. Performances on Defects4J \cite{dataset:Defects4j} and QuixBugs \cite{dataset:QuixBugs} are represented by the number of \textbf{correct/plausible } patches. The other three datasets (BFP(small) \cite{Tufano19}, CodRep \cite{dataset:CodRep}, \cite{TFix21}) only provide human-written patches for validation so the evaluation metric is the \textbf{top-k accuracy}.

Among all NPR approaches, Recoder \cite{Recoder21} creates the most complex encoder that receives the code text, AST and Tree Path as inputs all at once. It generated 71 correct patches for Defects4J (V1.2)'s 395 bugs. According to their original report, its performance is significantly better than that of other NPR systems as well as many test-suite-based approaches. On the java part of QuixBugs, Cure \cite{Cure21} produced 26 correct patches and 35 plausible patches among 40 bugs. It enhanced the model of CoCoNut \cite{CoCoNut20} with combining GPT \cite{GPT} that pre-trained on a large code corpus. On CodRep, SequenceR \cite{SequenceR19} achieves a double higher accuracy with class-level context and additional copy mechanism compared with the baseline in the top-50 candidates. TFix \cite{TFix21} that uses a pretrained T5 model \cite{T5} reported they achieve a fix rate of 46.30\% within top-5 candidates on the Restricted Text dataset compared with CoCoNut \cite{CoCoNut20} and SequenceR \cite{SequenceR19}. The small version of BFP only contains bug-fix pairs that are less than 50 tokens. On this dataset, Tang \cite{Tang21} perform better with grammar-based generation than vanilla sequence-to-sequence model \cite{Tufano19}. It has shown a top-1 accuracy of 11.47\%.

\section{Challenges and Solutions}\label{sec:challenge}
In this section, we will identify the challenges on designing an effective NPR systems and discuss existing solutions. We begin with an overview on typical challenges on each module of the NPR procedure in Section \ref{Overview}. Then we provide concrete discussion on corresponding solutions and sum up with general conclusions in Section \ref{OOV} to \ref{REd}.
\begin{table*}[htb]
  \centering
  \setlength{\abovecaptionskip}{0cm}
  \caption{Challenges and existing solutions for each module of the NPR system}
    \begin{tabular}{|l|p{8.584em}|p{25.71em}|}
    \hline
    \textbf{Module} & \multicolumn{1}{l|}{\textbf{Challenge}} & \multicolumn{1}{l|}{\textbf{Solution}} \bigstrut\\
    \hline
    \multirow{4}[8]{*}{Preprocessing} & Limit use of code-related information & Extract context arrounding the buggy code. The context can be neighbor lines \cite{TFix21}, buggy method \cite{Tufano19,DLFix20,CoCoNut20,hoppity20,Tang21,Cure21} or a buggy class \cite{SequenceR19}. \bigstrut\\
\cline{3-3}          & \multicolumn{1}{l|}{} & Construct features from AST \cite{codit20,DLFix20,Recoder21,Tang21,hoppity20} , or using a Static Analyzer \cite{TFix21}. \bigstrut\\
\cline{2-3}          & \multirow{2}[4]{*}{the OOV problem} & Simlify Identifiers \cite{Tufano19,Recoder21} or Literals \cite{Tufano19,CoCoNut20,Cure21}.  \bigstrut\\
\cline{3-3}          & \multicolumn{1}{r|}{} & Split infrequent words into frequent sub units with BPE \cite{transformer-edit20,Cure21,TFix21,CodeBert-finetune21} or Camel-aware tokenization \cite{CoCoNut20,Cure21}. \bigstrut\\
    \hline
    Input Representing & Limit use of code-related information & Utilizing tree-based encoders for AST-based inputs \cite{DLFix20,Recoder21,Tang21}, GNN for Graph-based inputs \cite{hoppity20}. \bigstrut\\
    \hline
    Output Searching & Large search space & Using Copy Mechanism \cite{codit20,SequenceR19,Recoder21} \bigstrut\\
    \hline
    Patch Ranking & Large search space & Perform a code-aware filter \cite{Cure21} or a DL-based classifier \cite{DLFix20}  \bigstrut\\
    \hline
    \end{tabular}%
  \label{tab:challenges&solutions}%
\end{table*}%
\subsection{Overview} \label{Overview}
Although some researchers \cite{CoCoNut20,Cure21,Recoder21} have report that they can achieve State-Of-The-Art (SOTA) results compared to test-suite-base APR, there is still great room for improvement. The obstacles come from differences between natural and programming languages, as well as that between the program repair and the translation task. These differences bring in challenges on different parts of approaches. We identify the main challenges as follows:

    \textbf{Out-Of-Vocabulary (OOV).} Programming languages use an open vocabulary, but DNN models use a pre-defined fix one. When dealing with tokens that are out of the predefined vocabulary, DNN's predictions will become imprecise. Since natural elements in the source code such as identifiers and literals are named subjectively by programmers, the OOV problem will be more troublesome when adopting neural models on the APR task.
    
    \textbf{Limited use of code-related information.} Programs written in high-level languages provide richer information than natural languages. For example, an Abstract-Syntax-Tree (AST) contains much more structural information compared with textual codes. Such code-related features can be used to help the model to be aware of more domain knowledge of the programming language. However, ordinary models in NLP field can not deal with these code-related information properly.
    
    \textbf{Large searching space.} Generally, a NPR model generates $V_s^l$ candidate patches for each bug where $V_s$ is the size of the vocabulary (usually at tens of thousands level) and $l$ is the length of the output. When generating patches, the NPR model choices one token from the vocabulary at each time. Obviously, such a huge search space will lead to a decrease on the model performance.

A summary of these challenges and corresponding solutions is described in Table \ref{tab:challenges&solutions}. We categorized them into four aforementioned phases of the overall NPR procedure. These phases face different challenges and necessitate different design choices to address them. First, in the \textit{preprocessing} phase, textual codes need to be divided into a list of tokens. All of these tokens together form a fixed vocabulary that defines the input space. This step is challenging since source codes may contain more OOV words, causing the model to make wrong predictions. Then in the \textit{input representing} phase, additional features constructed previously pose the challenge of designing specific neural encoders to better capture their semantics. Next, in the \textit{output searching} phase, the decoder module searches for potential fixes during the vocabulary-defined output space. It is challenging to select correct candidates from tens of thousands of choices. Finally, the generated candidate patches must be ranked and filtered in order to minimize validation costs. As a result, an efficient patch ranking strategy is required to significantly reduce the size of the candidate set while maintaining the correct patch hit rate. In the following part, we will show how previous studies solute these challenges and discuss the advantages and shortcomings of various solutions.

\subsection{Solutions to OOV} \label{OOV}
The first challenge is the OOV problem. When dealing with OOV words, the NPR model will gain a poor performance \cite{SequenceR19}. For the fact that programming languages use a more irregular vocabulary, the OOV problem will be even more serious when using Neural models to solve the APR task. In order to alleviate the OOV problem, researchers have taken several measurements reflected by design choices on the \textit{Code Abstraction} and the \textit{Code Tokenization} phase. The general purpose of existing solutions is to convert uncommon words into common ones. To this aim, some approaches choose to rename the natural elements in the source code such as identifiers \cite{Tufano19,Recoder21} and literals \cite{Tufano19,CoCoNut20,Cure21}. Another way is to divide word-piece tokens into sub units with BPE \cite{transformer-edit20,Cure21,TFix21,CodeBert-finetune21} or code-aware tokenization \cite{CoCoNut20,Cure21} that stems from commonly used naming rules of programming languages.
We look through their solutions on these two phases and further get general conclusions as follows:

\textbf{1. Abstraction of codes can reduce the size of the vocabulary, thus mitigating the OOV problem.} Code abstraction is a straightforward way that works on solving the OOV problem by renaming the natural elements in the source code to simplify the vocabulary. As an evidence, with abstract identifiers and literals, the accuracy of Tufano's model \cite{Tufano19} can be improved by 104\% compared to no-abstraction one on the CodRep \cite{dataset:CodRep} dataset. But such abstraction can also lead to a reduction of the applicability to real-world debugging scenarios. For example, Codit \cite{codit20} mentioned that the model trained with abstract identifiers can't handle the situation when the fix must introduce a new identifier not appeared in the source program. As we concluded, the best practice on code abstraction is to abstract all literals \cite{CoCoNut20} and uncommon identifiers \cite{Recoder21}, which can benefit a large performance improvement with a trivial decrease of applicability. 

\textbf{2. BPE-based tokenization may be not suitable for APR task.} BPE \cite{BPE} is a commonly used tokenization method in the NLP field. It can mitigate the OOV problem by splitting sentences into a series of sub-words. However, it is not suitable for the APR task for two reasons. Firstly, dividing the program into sub words will destroy the lexical and syntactic structure within the code text. Approaches that adopts BPE tokenization \cite{transformer-edit20,Cure21,CodeBert-finetune21,TFix21} can only incorporate textual features since structural features like AST need a word-piece tokenization. Secondly, a sentence toked by BPE will become much longer than one toked at word-level. The gain that BPE brings by solving OOV is less than the loss on a longer input, since DNN-based models has a weak ability to handle long inputs. According to the results of Tufano \cite{Tufano19} and SequenceR \cite{SequenceR19}, the accuracy of the same model on medium inputs (50 to 100 tokens) is 30\% - 60\% lower than that on small inputs (less than 50 tokens). We noticed that the existing BPE-based methods \cite{Cure21,TFix21} with good performance largely benefit from pre-train technologies which are too heavy and costly for APR task.
\subsection{Usage of code-related information}
Compared with natural languages, programming languages have more significant features. Proper use of these additional features can be beneficial to model's ability on understanding and fixing buggy programs. There are two core issues: how to construct features from source codes and how to represent features, which is reflected by the design choices on \textit{Preprocessing} and \textit{Input Representing} module. Extra features come from AST of the program \cite{codit20,DLFix20,Recoder21,Tang21,hoppity20} or a static program analyzer \cite{TFix21}. To extract AST-based features, tree-based \cite{DLFix20,Recoder21,Tang21} or graph-based \cite{hoppity20} are used in the \textit{Input Representing} phase. 

\textbf{1. The introduction of grammar rules is helpful for generating compilable patches.} Patches produced by vanilla encoder-decoder models may not be syntactically correct because they are generated purely based on probability. These patches would definitely be discarded because they can't even be compiled. Approaches without grammar constrains \cite{CoCoNut20,SequenceR19,Cure21} generate patches with an average compilable rate less than 40\%. Grammar-aware models \cite{codit20,Tang21} can learn to generate syntactically correct patches by representing grammar constraints with CFG rules, thus reaching a higher accuracy in the same-size candidate set. For one case, Codit \cite{codit20} exports an improvement with 63\% on suggesting correct patches within 5 candidates, compared with SequenceR \cite{SequenceR19}. However, such introduction of rules is not enough. Human programmers can also ensure correct syntax when writing code in text form. Instead of outputting CFG rules to form an AST, future NPR systems should ensure the syntax correctness when outputting textual codes.

\textbf{2. Structural models can be more precise at a price of decreasing the applicability.} There are two ways to represent AST: using structural models that suitable for tree-structure inputs or traversing AST to get sequential inputs. Most of the methods that use the former scheme \cite{DLFix20,hoppity20,Recoder21} obtained golden performance, benefiting from structural models' ability to represent AST. However, structural models have more restrictions on the form of input and output, which limits the repair scope of NPR systems. For example, DLFix \cite{DLFix20} that adopts a Tree-LSTM \cite{Tree-LSTM} based encoder-decoder model only works on one statement bugs and the fix and bug location have to be the same. Similarly, Recoder \cite{Recoder21} that uses a Tree-based decoder can only repair one-hunk bugs. As s comparison, the sequential encoder-decoder model can deal with any bug within the source input. We suggest that these two kinds of models can be applied to different scenarios. For example, for small or informal code programs, a sequential encoder-decoder can make a quick fix. When dealing with compilable programs in projects, a structural model can obtain a more precise prediction.

\subsection{Reduction of the search space} \label{REd}
Correct patches are sparse in the search space \cite{efficiency}. For NPR, large search space will have two negative effects. During the \textit{Output Searching} phase, it will decrease the performance since correct patches are sparse in the search space \cite{efficiency}. And during the \textit{Patch Ranking} phase, the large search space will result in too much patch candidates that are time-costly to validation. Therefore, a reduction strategy is necessary to make an efficient search. We review included studies' design choices at these two phases and sum up with two conclusions related to the search space.

\textbf{1. A copy generator is an efficient way to reduce the search space.} The copy mechanism \cite{copy} allows the model to select tokens from the source program as outputs during the decoding phase. In a sense, it enables the model to learn to keep correct tokens and only modify the buggy part. The ablation results of SequenceR \cite{SequenceR19} have demonstrated the remarkable improvement brought by copy mechanism. In practice, the copy mechanism can be easily integrated into various neural models \cite{codit20,transformer-edit20,SequenceR19,Recoder21}. 

\textbf{2. The number of candidates is not \textit{the more, the better}.}
Since the NPR system is a kind of probability model, a larger candidate number surely has a higher probability to contain the correct patch. However, the improvement on accuracy brought by more candidates will largely decline with the increase of the beam size according to existing studies \cite{Tufano19,codit20,DLFix20,transformer-edit20}. Such performance gain brought by a larger candidate number can result in a significant increase in time costs. For example, in Defects4J \cite{dataset:Defects4j}, a project may cost several minutes to compile and run the test cases. We observe that CURE \cite{Cure21} use a candidate number of 5,000. It means that using CURE \cite{Cure21} to fix a bug may cost several days (at the worst situation). Therefore, we suggest that follow-up researchers should also pay attention to the balance between performance and the time cost. 

\section{Conclusion and Future Directions}\label{sec:future}
This paper attempted to provide an elaborate overview on architectures, challenges and corresponding solutions of latest NPR Systems. To locate important design points, we proposed to explore the design space of each part of the NPR procedure. Furthermore, we identified three challenges current NPR approaches encountered and discussed the effect of existing solutions. In addition to better solutions on the challenges mentioned in this paper, we print out several promising future research directions on NPR: 

\textbf{More rules on generating.} From the standpoint of a developer, most patches generated by NPR models are of poor quality. Many NPR systems are unable to guarantee the syntax correctness of generated patches. The reason for this is that they are generated purely based on probability. In reality, however, the programming must adhere to certain syntax rules. The incorporation of these code-related specifications into NPR models will aid in the generation of more qualified programs.

\textbf{Explicable NPR models.} Since neural networks are black-box, there is low interpretability on how NPR models produce potential fixes. It may lead to users' distrust of the generated patches, resulting in limited applications of NPR models on real-world bug-fix scenarios. We suggested that besides the performance, efforts on making the model explainable should be dedicated. 

\textbf{Robustness measurement.} Existing NPR studies are all concerned with how many correct patches are generated in the experimental environment, while ignoring practical reliability metrics such as robustness. Due to some properties of neural networks, the NPR model may produce completely different predictions even when given inputs that are just slightly different. Such behaviors may result in errors that require a significant amount of manpower to trace and review. Therefore, metrics describing the robustness of the NPR model should be created. 
\bibliographystyle{ACM-Reference-Format}
\bibliography{Ref}


\begin{thebibliography}{62}


\ifx \showCODEN    \undefined \def \showCODEN     #1{\unskip}     \fi
\ifx \showDOI      \undefined \def \showDOI       #1{#1}\fi
\ifx \showISBNx    \undefined \def \showISBNx     #1{\unskip}     \fi
\ifx \showISBNxiii \undefined \def \showISBNxiii  #1{\unskip}     \fi
\ifx \showISSN     \undefined \def \showISSN      #1{\unskip}     \fi
\ifx \showLCCN     \undefined \def \showLCCN      #1{\unskip}     \fi
\ifx \shownote     \undefined \def \shownote      #1{#1}          \fi
\ifx \showarticletitle \undefined \def \showarticletitle #1{#1}   \fi
\ifx \showURL      \undefined \def \showURL       {\relax}        \fi
\providecommand\bibfield[2]{#2}
\providecommand\bibinfo[2]{#2}
\providecommand\natexlab[1]{#1}
\providecommand\showeprint[2][]{arXiv:#2}

\bibitem[Allamanis et~al\mbox{.}(2018)]%
        {GGNN}
\bibfield{author}{\bibinfo{person}{Miltiadis Allamanis}, \bibinfo{person}{Marc
  Brockschmidt}, {and} \bibinfo{person}{Mahmoud Khademi}.}
  \bibinfo{year}{2018}\natexlab{}.
\newblock \showarticletitle{Learning to Represent Programs with Graphs}. In
  \bibinfo{booktitle}{\emph{6th International Conference on Learning
  Representations, {ICLR} 2018, Vancouver, BC, Canada, April 30 - May 3, 2018,
  Conference Track Proceedings}}. \bibinfo{publisher}{OpenReview.net}.
\newblock
\urldef\tempurl%
\url{https://openreview.net/forum?id=BJOFETxR-}
\showURL{%
\tempurl}


\bibitem[Bahdanau et~al\mbox{.}(2015)]%
        {NMT-iclr15}
\bibfield{author}{\bibinfo{person}{Dzmitry Bahdanau},
  \bibinfo{person}{Kyunghyun Cho}, {and} \bibinfo{person}{Yoshua Bengio}.}
  \bibinfo{year}{2015}\natexlab{}.
\newblock \showarticletitle{Neural Machine Translation by Jointly Learning to
  Align and Translate}. In \bibinfo{booktitle}{\emph{3rd International
  Conference on Learning Representations, {ICLR} 2015, San Diego, CA, USA, May
  7-9, 2015, Conference Track Proceedings}},
  \bibfield{editor}{\bibinfo{person}{Yoshua Bengio} {and} \bibinfo{person}{Yann
  LeCun}} (Eds.).
\newblock
\urldef\tempurl%
\url{http://arxiv.org/abs/1409.0473}
\showURL{%
\tempurl}


\bibitem[Berabi et~al\mbox{.}(2021)]%
        {TFix21}
\bibfield{author}{\bibinfo{person}{Berkay Berabi}, \bibinfo{person}{Jingxuan
  He}, \bibinfo{person}{Veselin Raychev}, {and} \bibinfo{person}{Martin~T.
  Vechev}.} \bibinfo{year}{2021}\natexlab{}.
\newblock \showarticletitle{TFix: Learning to Fix Coding Errors with a
  Text-to-Text Transformer}. In \bibinfo{booktitle}{\emph{Proceedings of the
  38th International Conference on Machine Learning, {ICML} 2021, 18-24 July
  2021, Virtual Event}} \emph{(\bibinfo{series}{Proceedings of Machine Learning
  Research}, Vol.~\bibinfo{volume}{139})},
  \bibfield{editor}{\bibinfo{person}{Marina Meila} {and} \bibinfo{person}{Tong
  Zhang}} (Eds.). \bibinfo{publisher}{{PMLR}}, \bibinfo{pages}{780--791}.
\newblock
\urldef\tempurl%
\url{http://proceedings.mlr.press/v139/berabi21a.html}
\showURL{%
\tempurl}


\bibitem[Britton et~al\mbox{.}(2012)]%
        {quantify}
\bibfield{author}{\bibinfo{person}{Tom Britton}, \bibinfo{person}{Lisa Jeng},
  \bibinfo{person}{Graham Carver}, {and} \bibinfo{person}{Paul Cheak}.}
  \bibinfo{year}{2012}\natexlab{}.
\newblock \bibinfo{booktitle}{\emph{Quantify the time and cost saved using
  reversible debuggers}}.
\newblock \bibinfo{type}{{T}echnical {R}eport}. \bibinfo{institution}{Technical
  report, Cambridge Judge Business School}.
\newblock


\bibitem[Cao et~al\mbox{.}(2020)]%
        {survey-1}
\bibfield{author}{\bibinfo{person}{Heling Cao}, \bibinfo{person}{YangXia Meng},
  \bibinfo{person}{Jianshu Shi}, \bibinfo{person}{Lei Li},
  \bibinfo{person}{Tiaoli Liao}, {and} \bibinfo{person}{Chenyang Zhao}.}
  \bibinfo{year}{2020}\natexlab{}.
\newblock \showarticletitle{A Survey on Automatic Bug Fixing}. In
  \bibinfo{booktitle}{\emph{2020 6th International Symposium on System and
  Software Reliability (ISSSR)}}. \bibinfo{pages}{122--131}.
\newblock
\urldef\tempurl%
\url{https://doi.org/10.1109/ISSSR51244.2020.00029}
\showDOI{\tempurl}


\bibitem[Chakraborty et~al\mbox{.}(2020)]%
        {codit20}
\bibfield{author}{\bibinfo{person}{Saikat Chakraborty},
  \bibinfo{person}{Yangruibo Ding}, \bibinfo{person}{Miltiadis Allamanis},
  {and} \bibinfo{person}{Baishakhi Ray}.} \bibinfo{year}{2020}\natexlab{}.
\newblock \showarticletitle{Codit: Code editing with tree-based neural models}.
\newblock \bibinfo{journal}{\emph{IEEE Transactions on Software Engineering}}
  (\bibinfo{year}{2020}).
\newblock


\bibitem[Chen et~al\mbox{.}(2021c)]%
        {test-based:2021JAID}
\bibfield{author}{\bibinfo{person}{Liushan Chen}, \bibinfo{person}{Yu Pei},
  {and} \bibinfo{person}{Carlo~A. Furia}.} \bibinfo{year}{2021}\natexlab{c}.
\newblock \showarticletitle{Contract-Based Program Repair Without The
  Contracts: An Extended Study}.
\newblock \bibinfo{journal}{\emph{{IEEE} Trans. Software Eng.}}
  \bibinfo{volume}{47}, \bibinfo{number}{12} (\bibinfo{year}{2021}),
  \bibinfo{pages}{2841--2857}.
\newblock
\urldef\tempurl%
\url{https://doi.org/10.1109/TSE.2020.2970009}
\showDOI{\tempurl}


\bibitem[Chen et~al\mbox{.}(2021a)]%
        {transferlearning}
\bibfield{author}{\bibinfo{person}{Zimin Chen}, \bibinfo{person}{Steve
  Kommrusch}, {and} \bibinfo{person}{Martin Monperrus}.}
  \bibinfo{year}{2021}\natexlab{a}.
\newblock \showarticletitle{Neural Transfer Learning for Repairing Security
  Vulnerabilities in {C} Code}.
\newblock \bibinfo{journal}{\emph{CoRR}}  \bibinfo{volume}{abs/2104.08308}
  (\bibinfo{year}{2021}).
\newblock
\showeprint[arXiv]{2104.08308}
\urldef\tempurl%
\url{https://arxiv.org/abs/2104.08308}
\showURL{%
\tempurl}


\bibitem[Chen et~al\mbox{.}(2021b)]%
        {SequenceR19}
\bibfield{author}{\bibinfo{person}{Zimin Chen}, \bibinfo{person}{Steve
  Kommrusch}, \bibinfo{person}{Michele Tufano},
  \bibinfo{person}{Louis{-}No{\"{e}}l Pouchet}, \bibinfo{person}{Denys
  Poshyvanyk}, {and} \bibinfo{person}{Martin Monperrus}.}
  \bibinfo{year}{2021}\natexlab{b}.
\newblock \showarticletitle{SequenceR: Sequence-to-Sequence Learning for
  End-to-End Program Repair}.
\newblock \bibinfo{journal}{\emph{{IEEE} Trans. Software Eng.}}
  \bibinfo{volume}{47}, \bibinfo{number}{9} (\bibinfo{year}{2021}),
  \bibinfo{pages}{1943--1959}.
\newblock
\urldef\tempurl%
\url{https://doi.org/10.1109/TSE.2019.2940179}
\showDOI{\tempurl}


\bibitem[Chen and Monperrus(2018)]%
        {dataset:CodRep}
\bibfield{author}{\bibinfo{person}{Zimin Chen} {and} \bibinfo{person}{Martin
  Monperrus}.} \bibinfo{year}{2018}\natexlab{}.
\newblock \showarticletitle{The CodRep Machine Learning on Source Code
  Competition}.
\newblock \bibinfo{journal}{\emph{CoRR}}  \bibinfo{volume}{abs/1807.03200}
  (\bibinfo{year}{2018}).
\newblock
\showeprint[arXiv]{1807.03200}
\urldef\tempurl%
\url{http://arxiv.org/abs/1807.03200}
\showURL{%
\tempurl}


\bibitem[Cheng and Lapata(2016)]%
        {TextSummarization}
\bibfield{author}{\bibinfo{person}{Jianpeng Cheng} {and}
  \bibinfo{person}{Mirella Lapata}.} \bibinfo{year}{2016}\natexlab{}.
\newblock \showarticletitle{Neural Summarization by Extracting Sentences and
  Words}. In \bibinfo{booktitle}{\emph{Proceedings of the 54th Annual Meeting
  of the Association for Computational Linguistics, {ACL} 2016, August 7-12,
  2016, Berlin, Germany, Volume 1: Long Papers}}. \bibinfo{publisher}{The
  Association for Computer Linguistics}.
\newblock
\urldef\tempurl%
\url{https://doi.org/10.18653/v1/p16-1046}
\showDOI{\tempurl}


\bibitem[Devlin et~al\mbox{.}(2019)]%
        {pretrain-BERT}
\bibfield{author}{\bibinfo{person}{Jacob Devlin}, \bibinfo{person}{Ming{-}Wei
  Chang}, \bibinfo{person}{Kenton Lee}, {and} \bibinfo{person}{Kristina
  Toutanova}.} \bibinfo{year}{2019}\natexlab{}.
\newblock \showarticletitle{{BERT:} Pre-training of Deep Bidirectional
  Transformers for Language Understanding}. In
  \bibinfo{booktitle}{\emph{Proceedings of the 2019 Conference of the North
  American Chapter of the Association for Computational Linguistics: Human
  Language Technologies, {NAACL-HLT} 2019, Minneapolis, MN, USA, June 2-7,
  2019, Volume 1 (Long and Short Papers)}},
  \bibfield{editor}{\bibinfo{person}{Jill Burstein}, \bibinfo{person}{Christy
  Doran}, {and} \bibinfo{person}{Thamar Solorio}} (Eds.).
  \bibinfo{publisher}{Association for Computational Linguistics},
  \bibinfo{pages}{4171--4186}.
\newblock
\urldef\tempurl%
\url{https://doi.org/10.18653/v1/n19-1423}
\showDOI{\tempurl}


\bibitem[Dinella et~al\mbox{.}(2020)]%
        {hoppity20}
\bibfield{author}{\bibinfo{person}{Elizabeth Dinella}, \bibinfo{person}{Hanjun
  Dai}, \bibinfo{person}{Ziyang Li}, \bibinfo{person}{Mayur Naik},
  \bibinfo{person}{Le Song}, {and} \bibinfo{person}{Ke Wang}.}
  \bibinfo{year}{2020}\natexlab{}.
\newblock \showarticletitle{Hoppity: Learning Graph Transformations to Detect
  and Fix Bugs in Programs}. In \bibinfo{booktitle}{\emph{8th International
  Conference on Learning Representations, {ICLR} 2020, Addis Ababa, Ethiopia,
  April 26-30, 2020}}. \bibinfo{publisher}{OpenReview.net}.
\newblock
\urldef\tempurl%
\url{https://openreview.net/forum?id=SJeqs6EFvB}
\showURL{%
\tempurl}


\bibitem[Ding et~al\mbox{.}(2020)]%
        {transformer-edit20}
\bibfield{author}{\bibinfo{person}{Yangruibo Ding}, \bibinfo{person}{Baishakhi
  Ray}, \bibinfo{person}{Premkumar~T. Devanbu}, {and}
  \bibinfo{person}{Vincent~J. Hellendoorn}.} \bibinfo{year}{2020}\natexlab{}.
\newblock \showarticletitle{Patching as Translation: the Data and the
  Metaphor}. In \bibinfo{booktitle}{\emph{35th {IEEE/ACM} International
  Conference on Automated Software Engineering, {ASE} 2020, Melbourne,
  Australia, September 21-25, 2020}}. \bibinfo{publisher}{{IEEE}},
  \bibinfo{pages}{275--286}.
\newblock
\urldef\tempurl%
\url{https://doi.org/10.1145/3324884.3416587}
\showDOI{\tempurl}


\bibitem[Drain et~al\mbox{.}(2021)]%
        {DeepDebug}
\bibfield{author}{\bibinfo{person}{Dawn Drain}, \bibinfo{person}{Colin~B.
  Clement}, \bibinfo{person}{Guillermo Serrato}, {and} \bibinfo{person}{Neel
  Sundaresan}.} \bibinfo{year}{2021}\natexlab{}.
\newblock \showarticletitle{DeepDebug: Fixing Python Bugs Using Stack Traces,
  Backtranslation, and Code Skeletons}.
\newblock \bibinfo{journal}{\emph{CoRR}}  \bibinfo{volume}{abs/2105.09352}
  (\bibinfo{year}{2021}).
\newblock
\showeprint[arXiv]{2105.09352}
\urldef\tempurl%
\url{https://arxiv.org/abs/2105.09352}
\showURL{%
\tempurl}


\bibitem[Feng et~al\mbox{.}(2020)]%
        {CodeBERT}
\bibfield{author}{\bibinfo{person}{Zhangyin Feng}, \bibinfo{person}{Daya Guo},
  \bibinfo{person}{Duyu Tang}, \bibinfo{person}{Nan Duan},
  \bibinfo{person}{Xiaocheng Feng}, \bibinfo{person}{Ming Gong},
  \bibinfo{person}{Linjun Shou}, \bibinfo{person}{Bing Qin},
  \bibinfo{person}{Ting Liu}, \bibinfo{person}{Daxin Jiang}, {and}
  \bibinfo{person}{Ming Zhou}.} \bibinfo{year}{2020}\natexlab{}.
\newblock \showarticletitle{CodeBERT: {A} Pre-Trained Model for Programming and
  Natural Languages}. In \bibinfo{booktitle}{\emph{Findings of the Association
  for Computational Linguistics: {EMNLP} 2020, Online Event, 16-20 November
  2020}} \emph{(\bibinfo{series}{Findings of {ACL}},
  Vol.~\bibinfo{volume}{{EMNLP} 2020})},
  \bibfield{editor}{\bibinfo{person}{Trevor Cohn}, \bibinfo{person}{Yulan He},
  {and} \bibinfo{person}{Yang Liu}} (Eds.). \bibinfo{publisher}{Association for
  Computational Linguistics}, \bibinfo{pages}{1536--1547}.
\newblock
\urldef\tempurl%
\url{https://doi.org/10.18653/v1/2020.findings-emnlp.139}
\showDOI{\tempurl}


\bibitem[Goues et~al\mbox{.}(2012)]%
        {test-based:2011genprog}
\bibfield{author}{\bibinfo{person}{Claire~Le Goues}, \bibinfo{person}{ThanhVu
  Nguyen}, \bibinfo{person}{Stephanie Forrest}, {and} \bibinfo{person}{Westley
  Weimer}.} \bibinfo{year}{2012}\natexlab{}.
\newblock \showarticletitle{GenProg: {A} Generic Method for Automatic Software
  Repair}.
\newblock \bibinfo{journal}{\emph{{IEEE} Trans. Software Eng.}}
  \bibinfo{volume}{38}, \bibinfo{number}{1} (\bibinfo{year}{2012}),
  \bibinfo{pages}{54--72}.
\newblock
\urldef\tempurl%
\url{https://doi.org/10.1109/TSE.2011.104}
\showDOI{\tempurl}


\bibitem[Gupta et~al\mbox{.}(2017)]%
        {deepfix}
\bibfield{author}{\bibinfo{person}{Rahul Gupta}, \bibinfo{person}{Soham Pal},
  \bibinfo{person}{Aditya Kanade}, {and} \bibinfo{person}{Shirish~K. Shevade}.}
  \bibinfo{year}{2017}\natexlab{}.
\newblock \showarticletitle{DeepFix: Fixing Common {C} Language Errors by Deep
  Learning}. In \bibinfo{booktitle}{\emph{Proceedings of the Thirty-First
  {AAAI} Conference on Artificial Intelligence, February 4-9, 2017, San
  Francisco, California, {USA}}}, \bibfield{editor}{\bibinfo{person}{Satinder
  Singh} {and} \bibinfo{person}{Shaul Markovitch}} (Eds.).
  \bibinfo{publisher}{{AAAI} Press}, \bibinfo{pages}{1345--1351}.
\newblock
\urldef\tempurl%
\url{http://aaai.org/ocs/index.php/AAAI/AAAI17/paper/view/14603}
\showURL{%
\tempurl}


\bibitem[Hata et~al\mbox{.}(2018)]%
        {Ratchet}
\bibfield{author}{\bibinfo{person}{Hideaki Hata}, \bibinfo{person}{Emad
  Shihab}, {and} \bibinfo{person}{Graham Neubig}.}
  \bibinfo{year}{2018}\natexlab{}.
\newblock \showarticletitle{Learning to Generate Corrective Patches using
  Neural Machine Translation}.
\newblock \bibinfo{journal}{\emph{CoRR}}  \bibinfo{volume}{abs/1812.07170}
  (\bibinfo{year}{2018}).
\newblock
\showeprint[arXiv]{1812.07170}
\urldef\tempurl%
\url{http://arxiv.org/abs/1812.07170}
\showURL{%
\tempurl}


\bibitem[Hochreiter and Schmidhuber(1996)]%
        {LSTM96}
\bibfield{author}{\bibinfo{person}{Sepp Hochreiter} {and}
  \bibinfo{person}{J{\"{u}}rgen Schmidhuber}.} \bibinfo{year}{1996}\natexlab{}.
\newblock \showarticletitle{{LSTM} can Solve Hard Long Time Lag Problems}. In
  \bibinfo{booktitle}{\emph{Advances in Neural Information Processing Systems
  9, NIPS, Denver, CO, USA, December 2-5, 1996}},
  \bibfield{editor}{\bibinfo{person}{Michael Mozer},
  \bibinfo{person}{Michael~I. Jordan}, {and} \bibinfo{person}{Thomas Petsche}}
  (Eds.). \bibinfo{publisher}{{MIT} Press}, \bibinfo{pages}{473--479}.
\newblock
\urldef\tempurl%
\url{http://papers.nips.cc/paper/1215-lstm-can-solve-hard-long-time-lag-problems}
\showURL{%
\tempurl}


\bibitem[Jiang et~al\mbox{.}(2018)]%
        {test-based:2018SimFix}
\bibfield{author}{\bibinfo{person}{Jiajun Jiang}, \bibinfo{person}{Yingfei
  Xiong}, \bibinfo{person}{Hongyu Zhang}, \bibinfo{person}{Qing Gao}, {and}
  \bibinfo{person}{Xiangqun Chen}.} \bibinfo{year}{2018}\natexlab{}.
\newblock \showarticletitle{Shaping program repair space with existing patches
  and similar code}. In \bibinfo{booktitle}{\emph{Proceedings of the 27th {ACM}
  {SIGSOFT} International Symposium on Software Testing and Analysis, {ISSTA}
  2018, Amsterdam, The Netherlands, July 16-21, 2018}},
  \bibfield{editor}{\bibinfo{person}{Frank Tip} {and} \bibinfo{person}{Eric
  Bodden}} (Eds.). \bibinfo{publisher}{{ACM}}, \bibinfo{pages}{298--309}.
\newblock
\urldef\tempurl%
\url{https://doi.org/10.1145/3213846.3213871}
\showDOI{\tempurl}


\bibitem[Jiang et~al\mbox{.}(2021)]%
        {Cure21}
\bibfield{author}{\bibinfo{person}{Nan Jiang}, \bibinfo{person}{Thibaud
  Lutellier}, {and} \bibinfo{person}{Lin Tan}.}
  \bibinfo{year}{2021}\natexlab{}.
\newblock \showarticletitle{{CURE:} Code-Aware Neural Machine Translation for
  Automatic Program Repair}. In \bibinfo{booktitle}{\emph{43rd {IEEE/ACM}
  International Conference on Software Engineering, {ICSE} 2021, Madrid, Spain,
  22-30 May 2021}}. \bibinfo{publisher}{{IEEE}}, \bibinfo{pages}{1161--1173}.
\newblock
\urldef\tempurl%
\url{https://doi.org/10.1109/ICSE43902.2021.00107}
\showDOI{\tempurl}


\bibitem[Just et~al\mbox{.}(2014)]%
        {dataset:Defects4j}
\bibfield{author}{\bibinfo{person}{Ren{\'{e}} Just}, \bibinfo{person}{Darioush
  Jalali}, {and} \bibinfo{person}{Michael~D. Ernst}.}
  \bibinfo{year}{2014}\natexlab{}.
\newblock \showarticletitle{Defects4J: a database of existing faults to enable
  controlled testing studies for Java programs}. In
  \bibinfo{booktitle}{\emph{International Symposium on Software Testing and
  Analysis, {ISSTA} '14, San Jose, CA, {USA} - July 21 - 26, 2014}},
  \bibfield{editor}{\bibinfo{person}{Corina~S. Pasareanu} {and}
  \bibinfo{person}{Darko Marinov}} (Eds.). \bibinfo{publisher}{{ACM}},
  \bibinfo{pages}{437--440}.
\newblock
\urldef\tempurl%
\url{https://doi.org/10.1145/2610384.2628055}
\showDOI{\tempurl}


\bibitem[Knuth(1968)]%
        {CFG-rules}
\bibfield{author}{\bibinfo{person}{Donald~E. Knuth}.}
  \bibinfo{year}{1968}\natexlab{}.
\newblock \showarticletitle{Semantics of Context-Free Languages}.
\newblock \bibinfo{journal}{\emph{Math. Syst. Theory}} \bibinfo{volume}{2},
  \bibinfo{number}{2} (\bibinfo{year}{1968}), \bibinfo{pages}{127--145}.
\newblock
\urldef\tempurl%
\url{https://doi.org/10.1007/BF01692511}
\showDOI{\tempurl}


\bibitem[Li et~al\mbox{.}(2019)]%
        {source3}
\bibfield{author}{\bibinfo{person}{Daoyuan Li}, \bibinfo{person}{Li Li},
  \bibinfo{person}{Dongsun Kim}, \bibinfo{person}{Tegawend{\'{e}}~F.
  Bissyand{\'{e}}}, \bibinfo{person}{David Lo}, {and} \bibinfo{person}{Yves~Le
  Traon}.} \bibinfo{year}{2019}\natexlab{}.
\newblock \showarticletitle{Watch out for this commit! {A} study of influential
  software changes}.
\newblock \bibinfo{journal}{\emph{J. Softw. Evol. Process.}}
  \bibinfo{volume}{31}, \bibinfo{number}{12} (\bibinfo{year}{2019}).
\newblock
\urldef\tempurl%
\url{https://doi.org/10.1002/smr.2181}
\showDOI{\tempurl}


\bibitem[Li et~al\mbox{.}(2020)]%
        {DLFix20}
\bibfield{author}{\bibinfo{person}{Yi Li}, \bibinfo{person}{Shaohua Wang},
  {and} \bibinfo{person}{Tien~N. Nguyen}.} \bibinfo{year}{2020}\natexlab{}.
\newblock \showarticletitle{DLFix: context-based code transformation learning
  for automated program repair}. In \bibinfo{booktitle}{\emph{{ICSE} '20: 42nd
  International Conference on Software Engineering, Seoul, South Korea, 27 June
  - 19 July, 2020}}, \bibfield{editor}{\bibinfo{person}{Gregg Rothermel} {and}
  \bibinfo{person}{Doo{-}Hwan Bae}} (Eds.). \bibinfo{publisher}{{ACM}},
  \bibinfo{pages}{602--614}.
\newblock
\urldef\tempurl%
\url{https://doi.org/10.1145/3377811.3380345}
\showDOI{\tempurl}


\bibitem[Lin et~al\mbox{.}(2017)]%
        {dataset:QuixBugs}
\bibfield{author}{\bibinfo{person}{Derrick Lin}, \bibinfo{person}{James
  Koppel}, \bibinfo{person}{Angela Chen}, {and} \bibinfo{person}{Armando
  Solar{-}Lezama}.} \bibinfo{year}{2017}\natexlab{}.
\newblock \showarticletitle{QuixBugs: a multi-lingual program repair benchmark
  set based on the quixey challenge}. In \bibinfo{booktitle}{\emph{Proceedings
  Companion of the 2017 {ACM} {SIGPLAN} International Conference on Systems,
  Programming, Languages, and Applications: Software for Humanity, {SPLASH}
  2017, Vancouver, BC, Canada, October 23 - 27, 2017}},
  \bibfield{editor}{\bibinfo{person}{Gail~C. Murphy}} (Ed.).
  \bibinfo{publisher}{{ACM}}, \bibinfo{pages}{55--56}.
\newblock
\urldef\tempurl%
\url{https://doi.org/10.1145/3135932.3135941}
\showDOI{\tempurl}


\bibitem[Liu et~al\mbox{.}(2020)]%
        {efficiency}
\bibfield{author}{\bibinfo{person}{Kui Liu}, \bibinfo{person}{Shangwen Wang},
  \bibinfo{person}{Anil Koyuncu}, \bibinfo{person}{Kisub Kim},
  \bibinfo{person}{Tegawend{\'{e}}~F. Bissyand{\'{e}}},
  \bibinfo{person}{Dongsun Kim}, \bibinfo{person}{Peng Wu},
  \bibinfo{person}{Jacques Klein}, \bibinfo{person}{Xiaoguang Mao}, {and}
  \bibinfo{person}{Yves~Le Traon}.} \bibinfo{year}{2020}\natexlab{}.
\newblock \showarticletitle{On the efficiency of test suite based program
  repair: {A} Systematic Assessment of 16 Automated Repair Systems for Java
  Programs}. In \bibinfo{booktitle}{\emph{{ICSE} '20: 42nd International
  Conference on Software Engineering, Seoul, South Korea, 27 June - 19 July,
  2020}}, \bibfield{editor}{\bibinfo{person}{Gregg Rothermel} {and}
  \bibinfo{person}{Doo{-}Hwan Bae}} (Eds.). \bibinfo{publisher}{{ACM}},
  \bibinfo{pages}{615--627}.
\newblock
\urldef\tempurl%
\url{https://doi.org/10.1145/3377811.3380338}
\showDOI{\tempurl}


\bibitem[Lutellier et~al\mbox{.}(2020)]%
        {CoCoNut20}
\bibfield{author}{\bibinfo{person}{Thibaud Lutellier},
  \bibinfo{person}{Hung~Viet Pham}, \bibinfo{person}{Lawrence Pang},
  \bibinfo{person}{Yitong Li}, \bibinfo{person}{Moshi Wei}, {and}
  \bibinfo{person}{Lin Tan}.} \bibinfo{year}{2020}\natexlab{}.
\newblock \showarticletitle{CoCoNuT: combining context-aware neural translation
  models using ensemble for program repair}. In
  \bibinfo{booktitle}{\emph{{ISSTA} '20: 29th {ACM} {SIGSOFT} International
  Symposium on Software Testing and Analysis, Virtual Event, USA, July 18-22,
  2020}}, \bibfield{editor}{\bibinfo{person}{Sarfraz Khurshid} {and}
  \bibinfo{person}{Corina~S. Pasareanu}} (Eds.). \bibinfo{publisher}{{ACM}},
  \bibinfo{pages}{101--114}.
\newblock
\urldef\tempurl%
\url{https://doi.org/10.1145/3395363.3397369}
\showDOI{\tempurl}


\bibitem[Martinez and Monperrus(2016)]%
        {test-based:2016Astor}
\bibfield{author}{\bibinfo{person}{Matias Martinez} {and}
  \bibinfo{person}{Martin Monperrus}.} \bibinfo{year}{2016}\natexlab{}.
\newblock \showarticletitle{{ASTOR:} a program repair library for Java (demo)}.
  In \bibinfo{booktitle}{\emph{Proceedings of the 25th International Symposium
  on Software Testing and Analysis, {ISSTA} 2016, Saarbr{\"{u}}cken, Germany,
  July 18-20, 2016}}, \bibfield{editor}{\bibinfo{person}{Andreas Zeller} {and}
  \bibinfo{person}{Abhik Roychoudhury}} (Eds.). \bibinfo{publisher}{{ACM}},
  \bibinfo{pages}{441--444}.
\newblock
\urldef\tempurl%
\url{https://doi.org/10.1145/2931037.2948705}
\showDOI{\tempurl}


\bibitem[Martinez and Monperrus(2018)]%
        {test-based:2018Cardumen}
\bibfield{author}{\bibinfo{person}{Matias Martinez} {and}
  \bibinfo{person}{Martin Monperrus}.} \bibinfo{year}{2018}\natexlab{}.
\newblock \showarticletitle{Ultra-Large Repair Search Space with Automatically
  Mined Templates: The Cardumen Mode of Astor}. In
  \bibinfo{booktitle}{\emph{Search-Based Software Engineering - 10th
  International Symposium, {SSBSE} 2018, Montpellier, France, September 8-9,
  2018, Proceedings}} \emph{(\bibinfo{series}{Lecture Notes in Computer
  Science}, Vol.~\bibinfo{volume}{11036})},
  \bibfield{editor}{\bibinfo{person}{Thelma~Elita Colanzi} {and}
  \bibinfo{person}{Phil McMinn}} (Eds.). \bibinfo{publisher}{Springer},
  \bibinfo{pages}{65--86}.
\newblock
\urldef\tempurl%
\url{https://doi.org/10.1007/978-3-319-99241-9\_3}
\showDOI{\tempurl}


\bibitem[Mashhadi and Hemmati(2021)]%
        {CodeBert-finetune21}
\bibfield{author}{\bibinfo{person}{Ehsan Mashhadi} {and} \bibinfo{person}{Hadi
  Hemmati}.} \bibinfo{year}{2021}\natexlab{}.
\newblock \showarticletitle{Applying CodeBERT for Automated Program Repair of
  Java Simple Bugs}. In \bibinfo{booktitle}{\emph{18th {IEEE/ACM} International
  Conference on Mining Software Repositories, {MSR} 2021, Madrid, Spain, May
  17-19, 2021}}. \bibinfo{publisher}{{IEEE}}, \bibinfo{pages}{505--509}.
\newblock
\urldef\tempurl%
\url{https://doi.org/10.1109/MSR52588.2021.00063}
\showDOI{\tempurl}


\bibitem[Monperrus(2018)]%
        {APR}
\bibfield{author}{\bibinfo{person}{Martin Monperrus}.}
  \bibinfo{year}{2018}\natexlab{}.
\newblock \showarticletitle{Automatic Software Repair: {A} Bibliography}.
\newblock \bibinfo{journal}{\emph{{ACM} Comput. Surv.}} \bibinfo{volume}{51},
  \bibinfo{number}{1} (\bibinfo{year}{2018}), \bibinfo{pages}{17:1--17:24}.
\newblock
\urldef\tempurl%
\url{https://doi.org/10.1145/3105906}
\showDOI{\tempurl}


\bibitem[Monperrus(2020)]%
        {monperrus2020living}
\bibfield{author}{\bibinfo{person}{Martin Monperrus}.}
  \bibinfo{year}{2020}\natexlab{}.
\newblock \showarticletitle{The living review on automated program repair}.
\newblock  (\bibinfo{year}{2020}).
\newblock


\bibitem[Monperrus and Martinez(2012)]%
        {source2}
\bibfield{author}{\bibinfo{person}{Martin Monperrus} {and}
  \bibinfo{person}{Matias Martinez}.} \bibinfo{year}{2012}\natexlab{}.
\newblock \showarticletitle{CVS-Vintage: A Dataset of 14 CVS Repositories of
  Java Software}.
\newblock  (\bibinfo{date}{12} \bibinfo{year}{2012}).
\newblock


\bibitem[Moon and Okazaki(2021)]%
        {OOV}
\bibfield{author}{\bibinfo{person}{Sangwhan Moon} {and} \bibinfo{person}{Naoaki
  Okazaki}.} \bibinfo{year}{2021}\natexlab{}.
\newblock \showarticletitle{Effects and Mitigation of Out-of-vocabulary in
  Universal Language Models}.
\newblock \bibinfo{journal}{\emph{J. Inf. Process.}}  \bibinfo{volume}{29}
  (\bibinfo{year}{2021}), \bibinfo{pages}{490--503}.
\newblock
\urldef\tempurl%
\url{https://doi.org/10.2197/ipsjjip.29.490}
\showDOI{\tempurl}


\bibitem[Namavar et~al\mbox{.}(2021)]%
        {controlled-experiment}
\bibfield{author}{\bibinfo{person}{Marjane Namavar}, \bibinfo{person}{Noor
  Nashid}, {and} \bibinfo{person}{Ali Mesbah}.}
  \bibinfo{year}{2021}\natexlab{}.
\newblock \showarticletitle{A Controlled Experiment of Different Code
  Representations for Learning-Based Bug Repair}.
\newblock \bibinfo{journal}{\emph{CoRR}}  \bibinfo{volume}{abs/2110.14081}
  (\bibinfo{year}{2021}).
\newblock
\showeprint[arXiv]{2110.14081}
\urldef\tempurl%
\url{https://arxiv.org/abs/2110.14081}
\showURL{%
\tempurl}


\bibitem[Nguyen et~al\mbox{.}(2013)]%
        {test-based:2013Semfix}
\bibfield{author}{\bibinfo{person}{Hoang Duong~Thien Nguyen},
  \bibinfo{person}{Dawei Qi}, \bibinfo{person}{Abhik Roychoudhury}, {and}
  \bibinfo{person}{Satish Chandra}.} \bibinfo{year}{2013}\natexlab{}.
\newblock \showarticletitle{SemFix: program repair via semantic analysis}. In
  \bibinfo{booktitle}{\emph{35th International Conference on Software
  Engineering, {ICSE} '13, San Francisco, CA, USA, May 18-26, 2013}},
  \bibfield{editor}{\bibinfo{person}{David Notkin}, \bibinfo{person}{Betty
  H.~C. Cheng}, {and} \bibinfo{person}{Klaus Pohl}} (Eds.).
  \bibinfo{publisher}{{IEEE} Computer Society}, \bibinfo{pages}{772--781}.
\newblock
\urldef\tempurl%
\url{https://doi.org/10.1109/ICSE.2013.6606623}
\showDOI{\tempurl}


\bibitem[Radford et~al\mbox{.}(2018)]%
        {GPT}
\bibfield{author}{\bibinfo{person}{Alec Radford}, \bibinfo{person}{Karthik
  Narasimhan}, \bibinfo{person}{Tim Salimans}, {and} \bibinfo{person}{Ilya
  Sutskever}.} \bibinfo{year}{2018}\natexlab{}.
\newblock \showarticletitle{Improving language understanding by generative
  pre-training}.
\newblock  (\bibinfo{year}{2018}).
\newblock


\bibitem[Raffel et~al\mbox{.}(2020)]%
        {T5}
\bibfield{author}{\bibinfo{person}{Colin Raffel}, \bibinfo{person}{Noam
  Shazeer}, \bibinfo{person}{Adam Roberts}, \bibinfo{person}{Katherine Lee},
  \bibinfo{person}{Sharan Narang}, \bibinfo{person}{Michael Matena},
  \bibinfo{person}{Yanqi Zhou}, \bibinfo{person}{Wei Li}, {and}
  \bibinfo{person}{Peter~J. Liu}.} \bibinfo{year}{2020}\natexlab{}.
\newblock \showarticletitle{Exploring the Limits of Transfer Learning with a
  Unified Text-to-Text Transformer}.
\newblock \bibinfo{journal}{\emph{J. Mach. Learn. Res.}}  \bibinfo{volume}{21}
  (\bibinfo{year}{2020}), \bibinfo{pages}{140:1--140:67}.
\newblock
\urldef\tempurl%
\url{http://jmlr.org/papers/v21/20-074.html}
\showURL{%
\tempurl}


\bibitem[Scholtes et~al\mbox{.}(2016)]%
        {source4}
\bibfield{author}{\bibinfo{person}{Ingo Scholtes}, \bibinfo{person}{Pavlin
  Mavrodiev}, {and} \bibinfo{person}{Frank Schweitzer}.}
  \bibinfo{year}{2016}\natexlab{}.
\newblock \showarticletitle{From Aristotle to Ringelmann: a large-scale
  analysis of team productivity and coordination in Open Source Software
  projects}.
\newblock \bibinfo{journal}{\emph{Empir. Softw. Eng.}} \bibinfo{volume}{21},
  \bibinfo{number}{2} (\bibinfo{year}{2016}), \bibinfo{pages}{642--683}.
\newblock
\urldef\tempurl%
\url{https://doi.org/10.1007/s10664-015-9406-4}
\showDOI{\tempurl}


\bibitem[See et~al\mbox{.}(2017)]%
        {copy}
\bibfield{author}{\bibinfo{person}{Abigail See}, \bibinfo{person}{Peter~J.
  Liu}, {and} \bibinfo{person}{Christopher~D. Manning}.}
  \bibinfo{year}{2017}\natexlab{}.
\newblock \showarticletitle{Get To The Point: Summarization with
  Pointer-Generator Networks}. In \bibinfo{booktitle}{\emph{Proceedings of the
  55th Annual Meeting of the Association for Computational Linguistics, {ACL}
  2017, Vancouver, Canada, July 30 - August 4, Volume 1: Long Papers}},
  \bibfield{editor}{\bibinfo{person}{Regina Barzilay} {and}
  \bibinfo{person}{Min{-}Yen Kan}} (Eds.). \bibinfo{publisher}{Association for
  Computational Linguistics}, \bibinfo{pages}{1073--1083}.
\newblock
\urldef\tempurl%
\url{https://doi.org/10.18653/v1/P17-1099}
\showDOI{\tempurl}


\bibitem[Sennrich et~al\mbox{.}(2016)]%
        {BPE}
\bibfield{author}{\bibinfo{person}{Rico Sennrich}, \bibinfo{person}{Barry
  Haddow}, {and} \bibinfo{person}{Alexandra Birch}.}
  \bibinfo{year}{2016}\natexlab{}.
\newblock \showarticletitle{Neural Machine Translation of Rare Words with
  Subword Units}. In \bibinfo{booktitle}{\emph{Proceedings of the 54th Annual
  Meeting of the Association for Computational Linguistics, {ACL} 2016, August
  7-12, 2016, Berlin, Germany, Volume 1: Long Papers}}. \bibinfo{publisher}{The
  Association for Computer Linguistics}.
\newblock
\urldef\tempurl%
\url{https://doi.org/10.18653/v1/p16-1162}
\showDOI{\tempurl}


\bibitem[Sun et~al\mbox{.}(2020)]%
        {TreeGen}
\bibfield{author}{\bibinfo{person}{Zeyu Sun}, \bibinfo{person}{Qihao Zhu},
  \bibinfo{person}{Yingfei Xiong}, \bibinfo{person}{Yican Sun},
  \bibinfo{person}{Lili Mou}, {and} \bibinfo{person}{Lu Zhang}.}
  \bibinfo{year}{2020}\natexlab{}.
\newblock \showarticletitle{TreeGen: {A} Tree-Based Transformer Architecture
  for Code Generation}. In \bibinfo{booktitle}{\emph{The Thirty-Fourth {AAAI}
  Conference on Artificial Intelligence, {AAAI} 2020, The Thirty-Second
  Innovative Applications of Artificial Intelligence Conference, {IAAI} 2020,
  The Tenth {AAAI} Symposium on Educational Advances in Artificial
  Intelligence, {EAAI} 2020, New York, NY, USA, February 7-12, 2020}}.
  \bibinfo{publisher}{{AAAI} Press}, \bibinfo{pages}{8984--8991}.
\newblock
\urldef\tempurl%
\url{https://aaai.org/ojs/index.php/AAAI/article/view/6430}
\showURL{%
\tempurl}


\bibitem[Tai et~al\mbox{.}(2015)]%
        {Tree-LSTM}
\bibfield{author}{\bibinfo{person}{Kai~Sheng Tai}, \bibinfo{person}{Richard
  Socher}, {and} \bibinfo{person}{Christopher~D. Manning}.}
  \bibinfo{year}{2015}\natexlab{}.
\newblock \showarticletitle{Improved Semantic Representations From
  Tree-Structured Long Short-Term Memory Networks}. In
  \bibinfo{booktitle}{\emph{Proceedings of the 53rd Annual Meeting of the
  Association for Computational Linguistics and the 7th International Joint
  Conference on Natural Language Processing of the Asian Federation of Natural
  Language Processing, {ACL} 2015, July 26-31, 2015, Beijing, China, Volume 1:
  Long Papers}}. \bibinfo{publisher}{The Association for Computer Linguistics},
  \bibinfo{pages}{1556--1566}.
\newblock
\urldef\tempurl%
\url{https://doi.org/10.3115/v1/p15-1150}
\showDOI{\tempurl}


\bibitem[Tang et~al\mbox{.}(2021)]%
        {Tang21}
\bibfield{author}{\bibinfo{person}{Yu Tang}, \bibinfo{person}{Long Zhou},
  \bibinfo{person}{Ambrosio Blanco}, \bibinfo{person}{Shujie Liu},
  \bibinfo{person}{Furu Wei}, \bibinfo{person}{Ming Zhou}, {and}
  \bibinfo{person}{Muyun Yang}.} \bibinfo{year}{2021}\natexlab{}.
\newblock \showarticletitle{Grammar-Based Patches Generation for Automated
  Program Repair}. In \bibinfo{booktitle}{\emph{Findings of the Association for
  Computational Linguistics: {ACL/IJCNLP} 2021, Online Event, August 1-6,
  2021}} \emph{(\bibinfo{series}{Findings of {ACL}},
  Vol.~\bibinfo{volume}{{ACL/IJCNLP} 2021})},
  \bibfield{editor}{\bibinfo{person}{Chengqing Zong}, \bibinfo{person}{Fei
  Xia}, \bibinfo{person}{Wenjie Li}, {and} \bibinfo{person}{Roberto Navigli}}
  (Eds.). \bibinfo{publisher}{Association for Computational Linguistics},
  \bibinfo{pages}{1300--1305}.
\newblock
\urldef\tempurl%
\url{https://doi.org/10.18653/v1/2021.findings-acl.111}
\showDOI{\tempurl}


\bibitem[Tufano et~al\mbox{.}(2019)]%
        {Tufano19}
\bibfield{author}{\bibinfo{person}{Michele Tufano}, \bibinfo{person}{Cody
  Watson}, \bibinfo{person}{Gabriele Bavota}, \bibinfo{person}{Massimiliano~Di
  Penta}, \bibinfo{person}{Martin White}, {and} \bibinfo{person}{Denys
  Poshyvanyk}.} \bibinfo{year}{2019}\natexlab{}.
\newblock \showarticletitle{An Empirical Study on Learning Bug-Fixing Patches
  in the Wild via Neural Machine Translation}.
\newblock \bibinfo{journal}{\emph{{ACM} Trans. Softw. Eng. Methodol.}}
  \bibinfo{volume}{28}, \bibinfo{number}{4} (\bibinfo{year}{2019}),
  \bibinfo{pages}{19:1--19:29}.
\newblock
\urldef\tempurl%
\url{https://doi.org/10.1145/3340544}
\showDOI{\tempurl}


\bibitem[Vasic et~al\mbox{.}(2019)]%
        {joint}
\bibfield{author}{\bibinfo{person}{Marko Vasic}, \bibinfo{person}{Aditya
  Kanade}, \bibinfo{person}{Petros Maniatis}, \bibinfo{person}{David Bieber},
  {and} \bibinfo{person}{Rishabh Singh}.} \bibinfo{year}{2019}\natexlab{}.
\newblock \showarticletitle{Neural Program Repair by Jointly Learning to
  Localize and Repair}. In \bibinfo{booktitle}{\emph{7th International
  Conference on Learning Representations, {ICLR} 2019, New Orleans, LA, USA,
  May 6-9, 2019}}. \bibinfo{publisher}{OpenReview.net}.
\newblock
\urldef\tempurl%
\url{https://openreview.net/forum?id=ByloJ20qtm}
\showURL{%
\tempurl}


\bibitem[Vaswani et~al\mbox{.}(2017)]%
        {transformer}
\bibfield{author}{\bibinfo{person}{Ashish Vaswani}, \bibinfo{person}{Noam
  Shazeer}, \bibinfo{person}{Niki Parmar}, \bibinfo{person}{Jakob Uszkoreit},
  \bibinfo{person}{Llion Jones}, \bibinfo{person}{Aidan~N. Gomez},
  \bibinfo{person}{Lukasz Kaiser}, {and} \bibinfo{person}{Illia Polosukhin}.}
  \bibinfo{year}{2017}\natexlab{}.
\newblock \showarticletitle{Attention is All you Need}. In
  \bibinfo{booktitle}{\emph{Advances in Neural Information Processing Systems
  30: Annual Conference on Neural Information Processing Systems 2017, December
  4-9, 2017, Long Beach, CA, {USA}}},
  \bibfield{editor}{\bibinfo{person}{Isabelle Guyon}, \bibinfo{person}{Ulrike
  von Luxburg}, \bibinfo{person}{Samy Bengio}, \bibinfo{person}{Hanna~M.
  Wallach}, \bibinfo{person}{Rob Fergus}, \bibinfo{person}{S.~V.~N.
  Vishwanathan}, {and} \bibinfo{person}{Roman Garnett}} (Eds.).
  \bibinfo{pages}{5998--6008}.
\newblock
\urldef\tempurl%
\url{https://proceedings.neurips.cc/paper/2017/hash/3f5ee243547dee91fbd053c1c4a845aa-Abstract.html}
\showURL{%
\tempurl}


\bibitem[Wan et~al\mbox{.}(2018)]%
        {CodeSummarization}
\bibfield{author}{\bibinfo{person}{Yao Wan}, \bibinfo{person}{Zhou Zhao},
  \bibinfo{person}{Min Yang}, \bibinfo{person}{Guandong Xu},
  \bibinfo{person}{Haochao Ying}, \bibinfo{person}{Jian Wu}, {and}
  \bibinfo{person}{Philip~S. Yu}.} \bibinfo{year}{2018}\natexlab{}.
\newblock \showarticletitle{Improving automatic source code summarization via
  deep reinforcement learning}. In \bibinfo{booktitle}{\emph{Proceedings of the
  33rd {ACM/IEEE} International Conference on Automated Software Engineering,
  {ASE} 2018, Montpellier, France, September 3-7, 2018}},
  \bibfield{editor}{\bibinfo{person}{Marianne Huchard},
  \bibinfo{person}{Christian K{\"{a}}stner}, {and} \bibinfo{person}{Gordon
  Fraser}} (Eds.). \bibinfo{publisher}{{ACM}}, \bibinfo{pages}{397--407}.
\newblock
\urldef\tempurl%
\url{https://doi.org/10.1145/3238147.3238206}
\showDOI{\tempurl}


\bibitem[Wen et~al\mbox{.}(2018)]%
        {test-based:2018CapGen}
\bibfield{author}{\bibinfo{person}{Ming Wen}, \bibinfo{person}{Junjie Chen},
  \bibinfo{person}{Rongxin Wu}, \bibinfo{person}{Dan Hao}, {and}
  \bibinfo{person}{Shing{-}Chi Cheung}.} \bibinfo{year}{2018}\natexlab{}.
\newblock \showarticletitle{Context-aware patch generation for better automated
  program repair}. In \bibinfo{booktitle}{\emph{Proceedings of the 40th
  International Conference on Software Engineering, {ICSE} 2018, Gothenburg,
  Sweden, May 27 - June 03, 2018}}, \bibfield{editor}{\bibinfo{person}{Michel
  Chaudron}, \bibinfo{person}{Ivica Crnkovic}, \bibinfo{person}{Marsha
  Chechik}, {and} \bibinfo{person}{Mark Harman}} (Eds.).
  \bibinfo{publisher}{{ACM}}, \bibinfo{pages}{1--11}.
\newblock
\urldef\tempurl%
\url{https://doi.org/10.1145/3180155.3180233}
\showDOI{\tempurl}


\bibitem[Wong et~al\mbox{.}(2016)]%
        {survey:FL}
\bibfield{author}{\bibinfo{person}{W.~Eric Wong}, \bibinfo{person}{Ruizhi Gao},
  \bibinfo{person}{Yihao Li}, \bibinfo{person}{Rui Abreu}, {and}
  \bibinfo{person}{Franz Wotawa}.} \bibinfo{year}{2016}\natexlab{}.
\newblock \showarticletitle{A Survey on Software Fault Localization}.
\newblock \bibinfo{journal}{\emph{{IEEE} Trans. Software Eng.}}
  \bibinfo{volume}{42}, \bibinfo{number}{8} (\bibinfo{year}{2016}),
  \bibinfo{pages}{707--740}.
\newblock
\urldef\tempurl%
\url{https://doi.org/10.1109/TSE.2016.2521368}
\showDOI{\tempurl}


\bibitem[Xin and Reiss(2017)]%
        {test-based:2017ssFix}
\bibfield{author}{\bibinfo{person}{Qi Xin} {and} \bibinfo{person}{Steven~P.
  Reiss}.} \bibinfo{year}{2017}\natexlab{}.
\newblock \showarticletitle{Leveraging syntax-related code for automated
  program repair}. In \bibinfo{booktitle}{\emph{Proceedings of the 32nd
  {IEEE/ACM} International Conference on Automated Software Engineering, {ASE}
  2017, Urbana, IL, USA, October 30 - November 03, 2017}},
  \bibfield{editor}{\bibinfo{person}{Grigore Rosu},
  \bibinfo{person}{Massimiliano~Di Penta}, {and} \bibinfo{person}{Tien~N.
  Nguyen}} (Eds.). \bibinfo{publisher}{{IEEE} Computer Society},
  \bibinfo{pages}{660--670}.
\newblock
\urldef\tempurl%
\url{https://doi.org/10.1109/ASE.2017.8115676}
\showDOI{\tempurl}


\bibitem[Xiong et~al\mbox{.}(2017)]%
        {test-based:2017ACS}
\bibfield{author}{\bibinfo{person}{Yingfei Xiong}, \bibinfo{person}{Jie Wang},
  \bibinfo{person}{Runfa Yan}, \bibinfo{person}{Jiachen Zhang},
  \bibinfo{person}{Shi Han}, \bibinfo{person}{Gang Huang}, {and}
  \bibinfo{person}{Lu Zhang}.} \bibinfo{year}{2017}\natexlab{}.
\newblock \showarticletitle{Precise condition synthesis for program repair}. In
  \bibinfo{booktitle}{\emph{Proceedings of the 39th International Conference on
  Software Engineering, {ICSE} 2017, Buenos Aires, Argentina, May 20-28,
  2017}}, \bibfield{editor}{\bibinfo{person}{Sebasti{\'{a}}n Uchitel},
  \bibinfo{person}{Alessandro Orso}, {and} \bibinfo{person}{Martin~P.
  Robillard}} (Eds.). \bibinfo{publisher}{{IEEE} / {ACM}},
  \bibinfo{pages}{416--426}.
\newblock
\urldef\tempurl%
\url{https://doi.org/10.1109/ICSE.2017.45}
\showDOI{\tempurl}


\bibitem[Yasunaga and Liang(2020)]%
        {DrRepair}
\bibfield{author}{\bibinfo{person}{Michihiro Yasunaga} {and}
  \bibinfo{person}{Percy Liang}.} \bibinfo{year}{2020}\natexlab{}.
\newblock \showarticletitle{Graph-based, Self-Supervised Program Repair from
  Diagnostic Feedback}. In \bibinfo{booktitle}{\emph{Proceedings of the 37th
  International Conference on Machine Learning, {ICML} 2020, 13-18 July 2020,
  Virtual Event}} \emph{(\bibinfo{series}{Proceedings of Machine Learning
  Research}, Vol.~\bibinfo{volume}{119})}. \bibinfo{publisher}{{PMLR}},
  \bibinfo{pages}{10799--10808}.
\newblock
\urldef\tempurl%
\url{http://proceedings.mlr.press/v119/yasunaga20a.html}
\showURL{%
\tempurl}


\bibitem[Yasunaga and Liang(2021)]%
        {break-it-fix-it}
\bibfield{author}{\bibinfo{person}{Michihiro Yasunaga} {and}
  \bibinfo{person}{Percy Liang}.} \bibinfo{year}{2021}\natexlab{}.
\newblock \showarticletitle{Break-It-Fix-It: Unsupervised Learning for Program
  Repair}. In \bibinfo{booktitle}{\emph{Proceedings of the 38th International
  Conference on Machine Learning, {ICML} 2021, 18-24 July 2021, Virtual Event}}
  \emph{(\bibinfo{series}{Proceedings of Machine Learning Research},
  Vol.~\bibinfo{volume}{139})}, \bibfield{editor}{\bibinfo{person}{Marina
  Meila} {and} \bibinfo{person}{Tong Zhang}} (Eds.).
  \bibinfo{publisher}{{PMLR}}, \bibinfo{pages}{11941--11952}.
\newblock
\urldef\tempurl%
\url{http://proceedings.mlr.press/v139/yasunaga21a.html}
\showURL{%
\tempurl}


\bibitem[Ye et~al\mbox{.}(2021)]%
        {execution-based}
\bibfield{author}{\bibinfo{person}{He Ye}, \bibinfo{person}{Matias Martinez},
  {and} \bibinfo{person}{Martin Monperrus}.} \bibinfo{year}{2021}\natexlab{}.
\newblock \showarticletitle{Neural Program Repair with Execution-based
  Backpropagation}.
\newblock \bibinfo{journal}{\emph{CoRR}}  \bibinfo{volume}{abs/2105.04123}
  (\bibinfo{year}{2021}).
\newblock
\showeprint[arXiv]{2105.04123}
\urldef\tempurl%
\url{https://arxiv.org/abs/2105.04123}
\showURL{%
\tempurl}


\bibitem[Yoshida et~al\mbox{.}(2020)]%
        {phoenix}
\bibfield{author}{\bibinfo{person}{Hiroaki Yoshida}, \bibinfo{person}{Rohan
  Bavishi}, \bibinfo{person}{Keisuke Hotta}, \bibinfo{person}{Yusuke Nemoto},
  \bibinfo{person}{Mukul~R. Prasad}, {and} \bibinfo{person}{Shinji Kikuchi}.}
  \bibinfo{year}{2020}\natexlab{}.
\newblock \showarticletitle{Phoenix: a tool for automated data-driven synthesis
  of repairs for static analysis violations}. In
  \bibinfo{booktitle}{\emph{{ICSE} '20: 42nd International Conference on
  Software Engineering, Companion Volume, Seoul, South Korea, 27 June - 19
  July, 2020}}, \bibfield{editor}{\bibinfo{person}{Gregg Rothermel} {and}
  \bibinfo{person}{Doo{-}Hwan Bae}} (Eds.). \bibinfo{publisher}{{ACM}},
  \bibinfo{pages}{53--56}.
\newblock
\urldef\tempurl%
\url{https://doi.org/10.1145/3377812.3382150}
\showDOI{\tempurl}


\bibitem[Yuan and Banzhaf(2020)]%
        {test-based:2020ARJA}
\bibfield{author}{\bibinfo{person}{Yuan Yuan} {and} \bibinfo{person}{Wolfgang
  Banzhaf}.} \bibinfo{year}{2020}\natexlab{}.
\newblock \showarticletitle{{ARJA:} Automated Repair of Java Programs via
  Multi-Objective Genetic Programming}.
\newblock \bibinfo{journal}{\emph{{IEEE} Trans. Software Eng.}}
  \bibinfo{volume}{46}, \bibinfo{number}{10} (\bibinfo{year}{2020}),
  \bibinfo{pages}{1040--1067}.
\newblock
\urldef\tempurl%
\url{https://doi.org/10.1109/TSE.2018.2874648}
\showDOI{\tempurl}


\bibitem[Zhong and Su(2015)]%
        {source1}
\bibfield{author}{\bibinfo{person}{Hao Zhong} {and} \bibinfo{person}{Zhendong
  Su}.} \bibinfo{year}{2015}\natexlab{}.
\newblock \showarticletitle{An Empirical Study on Real Bug Fixes}. In
  \bibinfo{booktitle}{\emph{37th {IEEE/ACM} International Conference on
  Software Engineering, {ICSE} 2015, Florence, Italy, May 16-24, 2015, Volume
  1}}, \bibfield{editor}{\bibinfo{person}{Antonia Bertolino},
  \bibinfo{person}{Gerardo Canfora}, {and} \bibinfo{person}{Sebastian~G.
  Elbaum}} (Eds.). \bibinfo{publisher}{{IEEE} Computer Society},
  \bibinfo{pages}{913--923}.
\newblock
\urldef\tempurl%
\url{https://doi.org/10.1109/ICSE.2015.101}
\showDOI{\tempurl}


\bibitem[Zhou et~al\mbox{.}(2012)]%
        {source5}
\bibfield{author}{\bibinfo{person}{Jian Zhou}, \bibinfo{person}{Hongyu Zhang},
  {and} \bibinfo{person}{David Lo}.} \bibinfo{year}{2012}\natexlab{}.
\newblock \showarticletitle{Where should the bugs be fixed? More accurate
  information retrieval-based bug localization based on bug reports}. In
  \bibinfo{booktitle}{\emph{34th International Conference on Software
  Engineering, {ICSE} 2012, June 2-9, 2012, Zurich, Switzerland}},
  \bibfield{editor}{\bibinfo{person}{Martin Glinz}, \bibinfo{person}{Gail~C.
  Murphy}, {and} \bibinfo{person}{Mauro Pezz{\`{e}}}} (Eds.).
  \bibinfo{publisher}{{IEEE} Computer Society}, \bibinfo{pages}{14--24}.
\newblock
\urldef\tempurl%
\url{https://doi.org/10.1109/ICSE.2012.6227210}
\showDOI{\tempurl}


\bibitem[Zhu et~al\mbox{.}(2021)]%
        {Recoder21}
\bibfield{author}{\bibinfo{person}{Qihao Zhu}, \bibinfo{person}{Zeyu Sun},
  \bibinfo{person}{Yuan{-}an Xiao}, \bibinfo{person}{Wenjie Zhang},
  \bibinfo{person}{Kang Yuan}, \bibinfo{person}{Yingfei Xiong}, {and}
  \bibinfo{person}{Lu Zhang}.} \bibinfo{year}{2021}\natexlab{}.
\newblock \showarticletitle{A syntax-guided edit decoder for neural program
  repair}. In \bibinfo{booktitle}{\emph{{ESEC/FSE} '21: 29th {ACM} Joint
  European Software Engineering Conference and Symposium on the Foundations of
  Software Engineering, Athens, Greece, August 23-28, 2021}},
  \bibfield{editor}{\bibinfo{person}{Diomidis Spinellis},
  \bibinfo{person}{Georgios Gousios}, \bibinfo{person}{Marsha Chechik}, {and}
  \bibinfo{person}{Massimiliano~Di Penta}} (Eds.). \bibinfo{publisher}{{ACM}},
  \bibinfo{pages}{341--353}.
\newblock
\urldef\tempurl%
\url{https://doi.org/10.1145/3468264.3468544}
\showDOI{\tempurl}


\end{thebibliography}

\end{document}